**Towards in vivo g-ratio mapping using MRI: unifying myelin and diffusion imaging**

Siawoosh Mohammadi (1,2)*, Martina F. Callaghan (3)

(1) Department of Systems Neuroscience, University Medical Center Hamburg-Eppendorf, Hamburg, Germany

(2) Department of Neurophysics, Max Planck Institute for Human Cognitive and Brain Sciences, Leipzig, Germany

(3) Wellcome Centre for Human Neuroimaging, UCL Queen Square Institute of Neurology, University College London, UK

(*) corresponding author,

Department of Systems Neuroscience, University Medical Center Hamburg-Eppendorf, Martinistraße 52

20246 Hamburg

Germany

Phone: +49-40-7410-59859

Fax:   +49-40-7410-59955

E-Mail: siawooshm@googlemail.com

*Running header:* **Towards in vivo g-ratio mapping**

*Abstract*

The g-ratio, quantifying the comparative thickness of the myelin sheath encasing an axon, is a geometrical invariant that has high functional relevance because of its importance in determining neuronal conduction velocity. Advances in MRI data acquisition and signal modelling have put *in vivo* mapping of the g-ratio, across the entire white matter, within our reach. This capacity would greatly increase our knowledge of the nervous system: how it functions, and how it is impacted by disease. This is the second review on the topic of g-ratio mapping using MRI. As such, it summarizes the most recent developments in the field, while also providing methodological background pertinent to aggregate g-ratio weighted mapping, and discussing pitfalls associated with these approaches. Using simulations based on recently published data, this review demonstrates the relevance of the calibration step for three myelin-markers (macromolecular tissue volume, myelin water fraction, and bound pool fraction). It highlights the need to estimate both the slope and offset of the relationship between these MRI-based markers and the true myelin volume fraction if we are really to achieve the goal of precise, high sensitivity g-ratio mapping *in vivo*. Other challenges discussed in this review further evidence the need for gold standard measurements of human brain tissue from *ex vivo* histology. We conclude that the quest to find the most appropriate MRI biomarkers to enable *in vivo* g-ratio mapping is ongoing, with the potential of many novel techniques yet to be investigated.

# Table of Contents

# 1  *Introduction*

The g-ratio is a geometrical invariant of axons quantifying their degree of myelination relative to their cross-sectional size. Coupled with the axonal diameter, the g-ratio is a key determinate of neuronal conduction velocity (Rushton, 1951; Chomiak and Hu, 2009; Schmidt and Knösche, 2019). Signal transmission along different axonal fibres can be regulated and synchronised by varying the degree of myelination, and therefore the g-ratio, to optimize cognitive function, sensory integration and motor skills (Fields, 2015). As the central nervous system appears to communicate at physical limits to constrain metabolic demands (Salami et al., 2003; Hartline and Colman, 2007; Coggan et al., 2015), small deviations from the optimal g-ratio value (0.6-0.8, (Rushton, 1951; Chomiak and Hu, 2009)) may have strong functional impact. For example, histological investigation has shown that the cortical g-ratio is higher in patients with multiple sclerosis, probably because of the de- and re-myelination processes associated with the disease and its progression (Albert et al., 2007). To understand such processes and their functional implications, clinical research and diagnostics would benefit greatly from the capacity to measure the g-ratio of fibre pathways *in vivo*.

Until recently, information about the g-ratio distributions have only been accessible by invasive methods such as *ex vivo* electron microscopy (Hildebrand and Hahn, 1978), which restricted analyses to small numbers of axons and a limited number of brain regions or pathways. The g-ratio measured by such techniques is denoted the microscopic g-ratio because of the extremely fine spatial resolution that can be achieved. Clearly using MRI to investigate the g-ratio *in vivo* would be highly desirable as it could provide whole brain information on a voxel-wise basis. Stikov et al. proposed the methodology by which such a non-invasive MR-based "aggregate" g-ratio could be measured (Stikov et al., 2011, 2015), which we denote in this review interchangeably the "MR g-ratio" or "g-ratio mapping". The MR g-ratio framework measures the ensemble average of an underlying, unresolved, microstructural distribution of g-ratios. Making a strong assumption that the g-ratio is constant within a voxel, Stikov et al. demonstrated, via a geometrical plausibility argument (Stikov et al., 2011, 2015), that this aggregate MR g-ratio can be computed on a voxel-wise basis from the ratio of the myelin and axonal volume fractions (MVF and AVF respectively). Establishing this relation was important because both the MVF and AVF can be estimated by combining biophysical models (Alexander et al., 2019; Novikov et al., 2019) and quantitative MRI within a framework known as *in vivo* histology using MRI (Weiskopf et al., 2015).

The challenge for, and validity of, *in vivo* g-ratio mapping centres on how precisely and accurately the AVF and MVF can be measured with the chosen MRI techniques. Three years ago, Campbell et al. thoroughly reviewed the methods of g-ratio mapping and highlighted potential pitfalls (Campbell et al., 2018). A key take home message of their review was the introduction of the qualifying term "weighted" into the name MR g-ratio, i.e. aggregated g-ratio **weighted** mapping. They proposed this qualifier to acknowledge the impact that any miscalibration between the MR-based myelin proxy and the true MVF would have. Typically, *ex vivo* electron microscopy measures of the MVF act as the gold standard for methodological assessment and calibration.

Despite the challenges associated with accurate measurement and calibration of the MVF and AVF, many studies have exploited the potential of *in vivo* g-ratio weighted imaging for a variety of different applications (see Table 1 for full details). These have ranged from g-ratio mapping in infants (Melbourne et al., 2016) and children (Dean et al., 2016) to healthy adults (Mohammadi et al., 2015; Mancini et al., 2018; Berman et al., 2019; Drakesmith et al., 2019), during healthy aging (Cercignani et al., 2017; Berman et al.,

2018) and as a result of pathological change (Hagiwara et al., 2017; Hori et al., 2018; Kamagata et al., 2019; Yu et al., 2019).

Since the review by Campbell et al. (Campbell et al., 2018), awareness has increased regarding the fact that g-ratio mapping with MRI can be confounded by the degree of miscalibration of the MRI-based MVF proxy. Furthermore, new methodological studies have been published on g-ratio weighted mapping, e.g. to assess its repeatability (Duval et al., 2018; Ellerbrock and Mohammadi, 2018a), and the reproducibility when the particular proxies used for the AVF and MVF are varied (Ellerbrock and Mohammadi, 2018a). A series of validation studies have also been conducted by the Does lab (Kelm et al., 2016; West et al., 2018a, 2018b) based on extensive histological data coupled with *ex vivo* MRI. These studies have provided insight into the relationship between the MR g-ratio, and the various MVF measures through comparison with the current gold standard of electron microscopy. A particular strength of these studies has been the inherently large dynamic range of the MVF and g-ratio owing to the fact that hyper- and hypo-myelinating mouse models were used. This has enabled the validity and sensitivity of g-ratio mapping to be more thoroughly investigated.

In this review, we provide the background information necessary to understand the MRI methodologies that have been used to date to quantify the MVF and AVF (or fibre volume fraction, FVF) *in vivo* specifically in the context of g-ratio mapping. We seek to unify the nomenclature describing the various myelin and diffusion models. Moreover, we use the findings of the aforementioned methodological studies in simulation based experiments to further understand the impact on the MR g-ratio of currently used calibration methods. We examine the accuracy of the estimates based on the three myelin markers most commonly used for g-ratio mapping: the bound pool fraction derived, from quantitative magnetisation transfer methods, the macromolecular tissue volume derived from proton density mapping, and the myelin water fraction derived from myelin water imaging. Finally, we provide an outlook on emerging approaches and what will be required to make g-ratio mapping with MRI a viable clinical tool.

|  | Biomarkers | | Subjects or Participants | Remarks |
|---|---|---|---|---|
|  | Axonal or Fibre volume fraction (AVF or FVF) | Myelin volume fraction (MVF) |  | LA.x and LM.x refer to limitations pertinent, respectively, to the AVF or MVF measure used. |
| (Stikov et al., 2011) | DWI1 (DTI) | SPGR (qMT) | 5C | First model relating g-ratio to MVF and AVF. It assumed constant g-ratio in a voxel, and parallel axons. FA was also related to FVF assuming parallel fibres. LA.1, LM.1, LM.9 |
| (Stikov et al., 2015) | DWI2.5 (NODDI) | SPGR (qMT) | 1C; 1P; 1Mc; | Revised g-ratio model. In this model, the g-ratio is still assumed to be constant in a voxel but the model was extended to nonparallel axons. LA.3, LA.4, LM.1, LM.9 |
| (Mohammadi et al., 2015) | DWI1 (TFD) | MPM with multi-echo SPGR (MTsat) | 36C | First group study on g-ratio mapping using the MPM and DTI protocol as biomarkers for MVF and FVF. LA.2, LM.1, LM.2, LM.9 |
| (West et al., 2016) | - | - | 6M | Revised MR g-ratio model validated on volume fractions from electron microscopy, revealing that the MR g-ratio is an area-weighted average of the microscopic g-ratio. |
| (Melbourne et al., 2016) | DWI2 (NODDI) | 2D GRASE (MET2) | 37PI | The g-ratio of preterm infants scanned (27 and 58 weeks). LA.3, LA.4, LM.6, LM.7 |
| (Dean et al., 2016) | DWI2 (NOODI) | SPGR & bSSFP (mcDESPOT) | 18I | g-ratio index changes across childhood (3 months to 7.5 years of |
| | | | | age).LA.3, LA.4, LM.3, LM.9 |
| (Hagiwara et al., 2017) | DWI2 (NODDI) | SyMRI | 20P | g-ratio index in multiple sclerosis patients. MVF was estimated via the SyMRI model (Warntjes et al., 2016). LA.3, LA.4, LM.10 |
| (Duval et al., 2017) | DWI20 (CHARMED) | SPGR (MTV) | 9C | g-ratio index in human spinal cord. LA.6, LM.4, LM.1, LM.9 |
| (Cercignani et al., 2017) | DWI2.4 (NODDI) | bSSFP (qMT) | 38C | qMT was calculated via in-house software. B1+ correction was not reported. Change of g-ratio as a function of age. LA.3,LA.4, LM.1, LM.9 |
| (Ellerbrock and Mohammadi, 2018a) | DWI1 (TFD), DWI2 (NODDI) | MPM with multi-echo SPGR (MTsat, MTV) | 12C, 10C | Four different g-ratio index maps were compared in a scan-rescan experiment between two groups of subjects (12 and 10 subjects). LA.2, LA.3, LA.4, LM.1, LM.2, LM.9 |
| (Berman et al., 2018) | DWI1 (DTI) | SPGR (MTV) | 92C; M15* | Change of g-ratio as a function of age. LA.1, LM.1, LM.4, LM.9 |
| (Duval et al., 2018) | As in (Stikov et al., 2011) | SPGR (MTV) | 8C | Scan-rescan MRI on spinal cord. LA.1, LM.1, LM.4, LM.9 |
| (Hori et al., 2018) | DWI1 (NODDI) | MPM with multi-echo SPGR (MTsat) | 24P | Clinical study: G-ratio maps of the spinal cord in Cervical Spondylotic Myelopathy. LA.3, LA.4, LM.2 , LM.9 |
| (Jung et al., 2018) | DWI2 (NODDI) | Multi-echo SPGR (MET2*) | 5C; 15M* | Two calibration methods for estimating MVF from myelin-water fraction. LA.3, LA.4, LM.1, LM.6, LM.8 |
| (Mancini et al., 2018) | DWI2.4 (1), DWI2.9 (2) (NODDI) | bSSFP (1), SPGR (2) (qMT) | 16C,15C | Same as in Cercignani et al., 2017. Two datasets, dataset one acquired at 1.5T (1) and dataset two (2) at 3T, each on a different imaging site. G-ratio used to introduce axonal myelination in connectomics. B1+ |

| | | | | correction was not reported for site (1). LA.3, LA.4, LM.1, LM.9 |
|---|---|---|---|---|
| (West et al., 2018a) | DWI6 (NODDI, WMTI, mcSMT) | 3D MSE (MET2) | 15M | Electron microscopy and ex vivo MRI of mice models with varying degree of myelination using multi-shell diffusion MRI and 3D spin echo sequence. LA.3, LA.4, LA.5, LM.1, LM.6, LM.8 |
| (Kamagata et al., 2019) | DWI2 (NODDI) | MPM with multi-echo SPGR (MTsat) | 14C;14P | The brain network topology was assessed using g-ratio as a marker for the connectivity strength, comparison between healthy controls and multiple sclerosis patients. LA.3, LA.4, LM.1, LM.2, LM.9 |
| (Yu et al., 2019) | DWI17.8 (3CM) | SPGR (MTV) | 19C; 30P | g-ratio and axon diameter mapping in patients with multiple sclerosis and healthy controls. LA.6, , LM.4, LM.9 |
| (Berman et al., 2019) | DWI1 (DTI) | SPGR (MTV) | 37C | Estimating conduction velocities in fibre pathways using g-ratio and tractography. Data: 37 subjects (20 younger and 17 older humans). LM.1, LM.4,LM.9 |
| (Drakesmith et al., 2019) | DWI6 (CHARMED) | SPGR & SSFP (mcDESPOT) | 21C | Estimating conduction velocities in the corpus callosum using g-ratio and axon diameters. LA.6, LM.3,LM.9 |
| TABLE 1: Summary of in vivo MR g-ratio mapping studies. Limitations associated with the biomarkers for MVF (LM) and AVF/FVF (LA) are summarized in Table A1. C = health human controls; I = infants; M = mice; Mc = macaque; P = human patients; PI = preterm infants; 3CM = ActiveAx-like model (Alexander et al., 2010); * The mice data from (West et al., 2018a) were used. |||||

## 2 Methodology

Biological tissue is formed of multiple microenvironments, which we refer to as compartments or pools. From an MRI perspective, key compartments in an imaging voxel comprised of human brain tissue, are those formed of aqueous and non-aqueous protons (Fig. 1a). The aqueous protons ($f_W$) appear in a variety of microenvironments including water trapped within the myelin sheaths of fibre pathways ($f_{MW}$), or contained within the intra- ($f_{AW}$) and extra-cellular spaces ($f_{EW}$), and cerebrospinal fluid ($f_{CSF}$). The non-aqueous protons are bound to macromolecules ($f_B$), including myelin ($f_{BM}$) and other macromolecules such as proteins ($f_{BNM}$). We express these compartments as fractions of the imaging voxel under the simplifying assumption that, while the relative contribution will spatially vary, every voxel is fully described by its content of water and bound protons, i.e. $f_W + f_B = 1$. Of these tissue compartments, it is the axonal and myelin-associated compartments that are important in the context of *in vivo* g-ratio mapping (section 2.1). With MRI we tailor our experiments to maximise our sensitivity to specific compartments with the aim of quantifying the MVF and AVF respectively. To date, g-ratio mapping studies have either used relaxometry (Fig. 1b) or magnetisation transfer (Fig. 1c) techniques to quantify the myelin compartment (section 2.2), while diffusion imaging has been used to quantify the axonal compartment (Fig. 1d and section 2.3). These different imaging modalities have each evolved specific nomenclature over the course of their development. In this review, we aim, wherever possible, to unify these disparate notations using the fractional contributions outlined above and illustrated in figure 1.

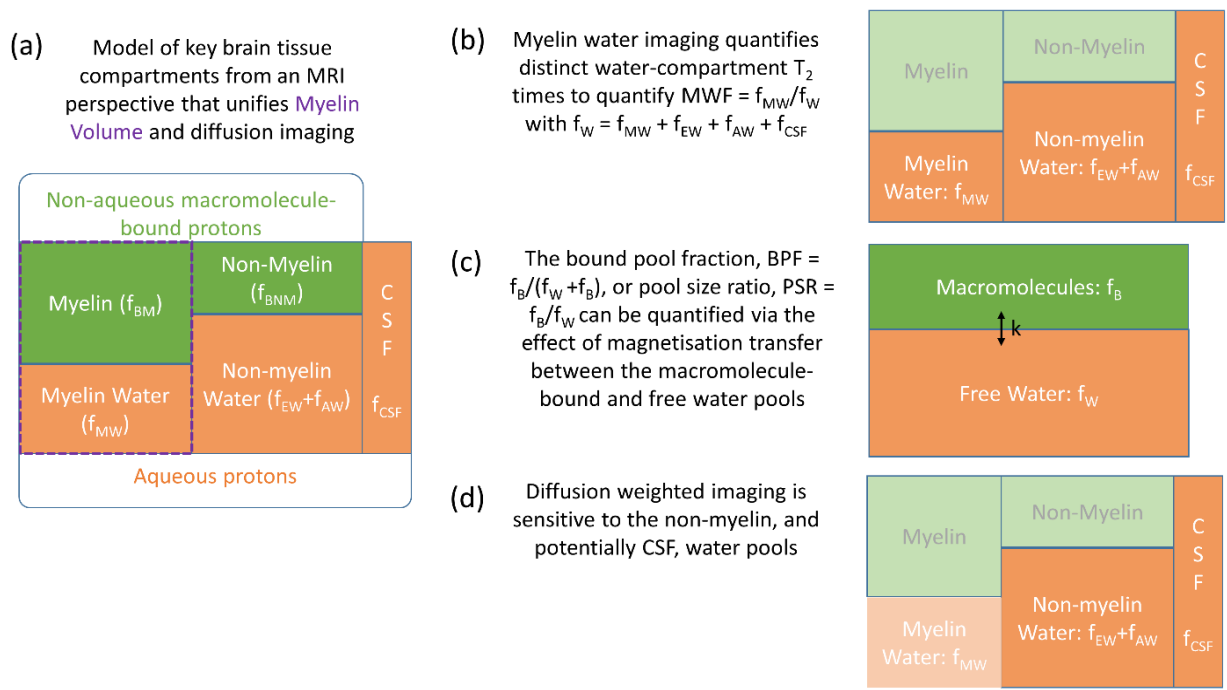

**Figure 1:** Unified model of myelin and diffusion imaging. To facilitate modelling, brain tissue is decomposed into four distinct tissue compartments (and CSF) that are of key relevance from an MRI perspective. These cover two broad categories: non-aqueous macromolecule-bound protons ($f_B$) and aqueous ($f_W$) protons, each of which may ($f_{MW}$, $f_{BM}$) or may not ($f_{AW}$, $f_{EW}$, $f_{CSF}$, $f_{BNM}$) be associated with myelin (a). Myelin water imaging specifically focuses on characterising the distinct water micro-environments to quantify the myelin water fraction, MWF, i.e. the fractional contribution from myelin-associated water, $f_{MW}$, relative to all water (b). Magnetisation transfer approaches focus instead on distinct macromolecular-bound and free water compartments, which can exchange magnetisation to quantify the bound pool fraction (BPF), i.e. the fractional contribution from the macromolecular environment (BPF=$f_B/(f_B+f_W)$, c). The diffusion weighted signal is sensitive to intra-axonal and extra-axonal water compartments, and potentially to an isotropic diffusion compartment such as CSF. By decomposing the signal, the intra-axonal water fraction (AWF) can be isolated: i.e. AWF = $f_{AW}/(f_{AW}+f_{EW}+f_{CSF})$.

## 2.1 The Aggregate g-ratio Model

Assuming a circular cross-section of axons, the microscopic $g$-ratio of an individual axon indexed by $j$ is defined as $g_j = \frac{RI}{RO}$, where $RI$ and $RO$ are the inner and outer radii of the fibre respectively (see Fig. 2a). All further considerations are targeting the white matter (WM), which is considered to be composed of three discrete, non-overlapping compartments: axonal (A), myelin (M), and extracellular (E). In this case, any sample volume of WM can be described by the volume fractions (VF) of each compartment, which sum up to one, i.e.: $AVF + MVF + EVF = 1$. Using this WM model, Stikov and colleagues (Stikov et al., 2011, 2015) suggested that the aggregated $g$-ratio in an MRI volume (Fig. 2b) can also be defined in terms of volume fractions as:

$$(1) \quad g_{MRI} = \sqrt{1 - \frac{MVF}{MVF + AVF}}$$

To derive the relationship in Eq. (1) (see also (Stikov et al., 2011, 2015)), the $g$-ratio in an MRI voxel is assumed to be constant (Fig. 2c), whereas there is no restriction on the orientation of the axons in the voxel (Fig. 2d). Shortly after the g-ratio model was introduced, (West et al., 2016) suggested that $g_{MRI}$ is in fact capturing the fibre-area-weighted mean (Fig. 2e) of all the microscopic g-ratios in the voxel (Fig. 2f). If the assumptions of Eq. (1) hold, this model can also be used with other imaging modalities (e.g., electron microscopy, where the $MVF$ and $AVF$ have been measured after segmentation of the image (West et al., 2016)). This efficient process allows the microscopic information obtained by these other modalities to be summarised over a spatial scale comparable to an MRI voxel, and therefore to be compared directly with the MR-based g-ratio in validation studies. The aggregate g-ratio model has been developed specifically for white matter (Stikov et al., 2011, 2015; Campbell et al., 2018), where biomarkers of the MVF and AVF can be measured with MRI. In the following sections we will first outline the methods that have been used to date to quantify MVF and AVF in the context of g-ratio mapping.

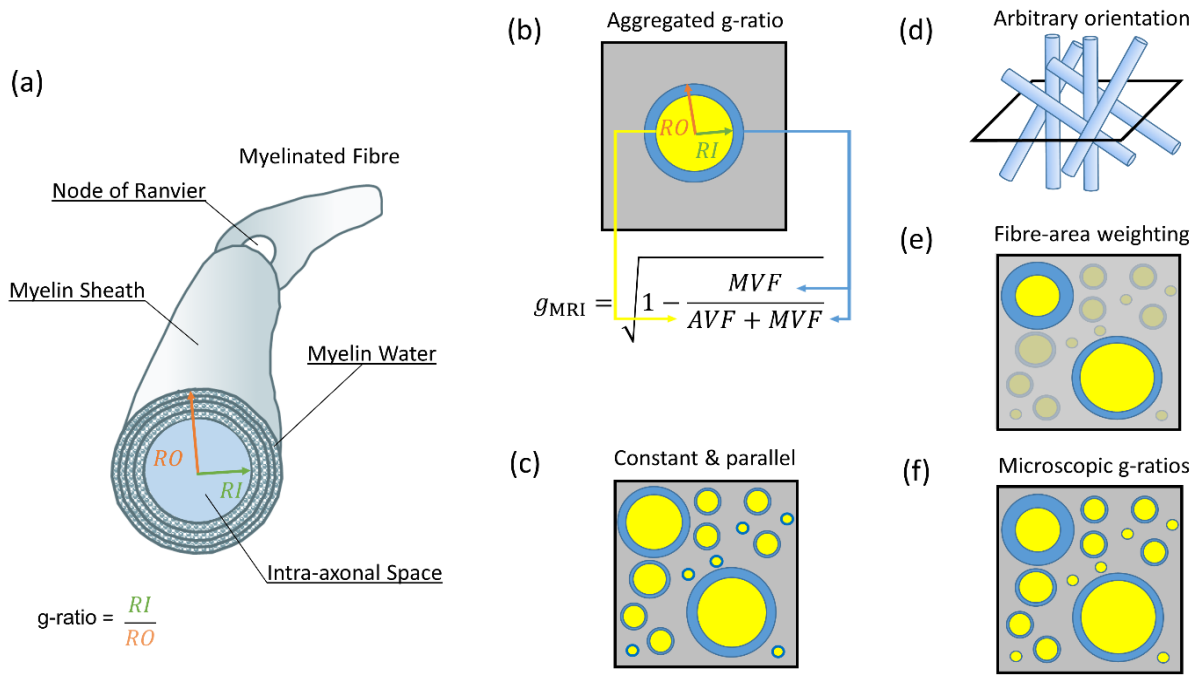

**Figure 2:** Schematic summary of the aggregated g-ratio model and its relation to the microscopic g-ratios. Myelinated axons (a) are represented by annual cylinders with myelin (blue) and axonal (yellow) compartments (b-f), other microstructural compartments are agglomerated in the background (grey). The aggregated g-ratio ($g_{MRI}$) can be formulated as a function of the axonal and myelin volume fractions ($AVF$ and $MVF$ respectively, b). In this model, all axons within a voxel are assumed to have the same g-ratio. In the initial model suggested by Stikov et al. in 2011, the axons were also assumed to be orientated in parallel (c). This assumption was subsequently relaxed (Stikov et al., 2015), allowing arbitrary axonal orientation (d). West et al. (2016) showed that the aggregated g-ratio is related to the fibre area-weighted mean of the microscopic g-ratios (f) – in the figure the weights are represented by the degree of transparency to indicate the weighting towards larger fibres (e).

*2.2    Myelin Volume Fraction*

A variety of different MRI-based measures have been used to characterise the myelin content within a voxel (Alonso- Ortiz et al., 2015; MacKay and Laule, 2016; Sled, 2018). Here we focus on myelin-water imaging (MWI) and magnetization transfer (MT) imaging. In both cases, each of which will be discussed in turn, the measure aims to be reflective of the fractional myelin content within the imaging volume, i.e. the MVF. This is done by quantifying either the myelin *water* fraction ($MWF = \frac{f_{MW}}{f_W}$, Fig. 1b) or the bound pool fraction ($BPF = \frac{f_B}{f_W + f_B}$, Fig. 1c). In either case, an additional calibration step is clearly required to convert the measure to the MVF ($\frac{f_{MW} + f_{BM}}{f_W + f_B}$) in order to accurately compute the g-ratio (West et al., 2018b). As noted by Campbell et al. (Campbell et al., 2018) this calibration step is crucial to the accuracy and precision of g-ratio mapping and will be discussed in detail in section 3.

*2.2.1  MWF based on Myelin Water Imaging*

Starting from Figure 1, the simplest water imaging model quantifies the density of free water protons within an imaging voxel, i.e. the proton density (PD) (Tofts, 2004). Under an assumption of complete longitudinal recovery within each repetition time, TR, the extrapolated MR signal at an echo time, TE, of 0ms ($S_0$) is proportional to the product of the fractional water content, $f_W$, a calibration factor, $C$, that accounts for the concentration of protons in the voxel relative to that of free water, and the spatially-varying receive field sensitivity, $R$, : $S_0 = R \, C \, f_W$ such that $f_W + f_B = 1$ (Fig. 1c). The receive field modulation must be estimated and removed ($S_0' = C \, f_W$, see section 2.2.3) prior to final calibration, which is done with respect to a reference, e.g. cerebrospinal fluid, CSF: $PD = \frac{S_0'}{S_{0,CSF}'} = f_W$.

This is equivalent to assuming that the volume fraction of macromolecules is zero ($f_B \approx 0$ and $f_W = 1$), i.e. $S'_{0,CSF} = C$. The remaining contents of the voxel have recently been referred to as the macromolecular tissue volume ($MTV = 1 - PD = f_B$) (Mezer et al., 2013). PD mapping typically makes no distinction between different water microenvironments (e.g. myelin water v's non-myelin water) and instead estimates the sum of contributions from all compartments (Fig. 1b,c) under the assumption of a mono-exponential signal decay. Therefore, MTV might vary with the minimum echo time, as well the echo spacing, at which the signal was sampled (more details can be found in (Tofts, 2004)).

By contrast, myelin water imaging (MWI, (Alonso- Ortiz et al., 2015)) extends this model to encompass multiple distinct water compartments, each with specific relaxation behaviour contingent on the local microenvironment. MWI quantifies myelin-associated aqueous protons in a voxel as a fraction of the total MR visible water signal, i.e. $MWF = \frac{f_{MW}}{f_W}$ as defined in Figure 2b. To date, three main approaches to myelin water imaging have been used for g-ratio mapping using MRI. Each technique exploits a different relaxation property to stratify the different tissue water compartments (MacKay and Laule, 2016): (1) multi-echo spin echo imaging to quantify compartment-specific transverse relaxation times (Melbourne et al., 2016; West et al., 2018a), $T_2$, (2) multi-echo gradient echo imaging to quantify compartment-specific effective transverse relaxation times (Jung et al., 2018) , $T_2^*$, and (3) multi-compartment driven equilibrium single pulse observation of T1 and T2 (mcDESPOT, (Deoni et al., 2008; Dean et al., 2016; Drakesmith et al., 2019)) to distinguish fast and slow relaxing compartments based on their distinct $T_1$ and $T_2$ relaxation and exchange behaviour.

In MWI, the MWF is most commonly estimated by characterising the proportion of the water signal originating from different microstructural environments based on their distinct transverse relaxation times ($T_2$). To do this, it is assumed that the residency time, $\tau$, of the protons in each water pool is sufficiently long that their distinct relaxation behaviour can be discerned. The case $\tau >> T_2$ indicates a slow exchange regime, which can equivalently be described by an exchange rate, $k = 1/\tau << 1/T_2$. In this case, multi-exponential behaviour, with a component originating from each of the water pools having distinct amplitude and relaxation times, can be discerned (Zimmerman and Brittin, 1957). Indeed, $T_2$ distributions from normal brain have been shown to contain multiple peaks that can be attributed to myelin water trapped between the lipid bilayers, intra/extracellular water and cerebral spinal fluid (Whittall et al., 1997; MacKay and Laule, 2016).

To quantify distinct $T_2$ times, data are typically acquired using a multi-echo spin echo readout with a range of echo times. Each voxel is assumed to contain contributions from an unspecified number of slow or non-exchanging environments, each with distinct $T_2$ decay times. Fitting the data to this model is typically done with a regularised non-negative least squares approach (Whittall and MacKay, 1989; MacKay et al., 2006), in which the regularisation ensures smoothly varying signal amplitudes as a function of $T_2$. After fitting, the myelin compartment is assigned to the short $T_2$ peaks, requiring a threshold $T_2$ time to be specified. The MWF is then estimated as the area under the peaks below this threshold $T_2$ time relative to the area under all peaks, i.e. $\frac{f_{MW}}{f_W}$ (MacKay et al., 2006). This ignores any differential weighting that might be present, for example due to compartment-specific $T_1$ times. In white matter, at least two different $T_2$ relaxation times have been reported, which are associated with different tissue compartments (MacKay et al., 2006; Cercignani et al., 2018): (1) myelin water having a $T_2$ of about 15 – 30 ms, and (2) water in the intra- and extra cellular spaces with a $T_2$ of about 80 – 90 ms, at 3T. It should also be noted that the $T_2$ relaxation times of the intra- and extra cellular spaces likely differ (Dortch et al., 2013; Veraart et al., 2018; McKinnon and Jensen, 2019) and that there is exchange between these compartments that also influences the $T_2$ distribution in white matter (Sled et al., 2004). These effects will be revisited in section 3.1.1 but have also been discussed in detail elsewhere (Does, 2018).

A similar approach uses a multi-echo gradient echo acquisition in lieu of acquiring spin echoes. In this case compartment-specific $T_2^*$ times are estimated instead of $T_2$ (Lenz et al., 2012; Sati et al., 2013). This approach is more time efficient, less vulnerable to transmit field inhomogeneity and less demanding from an RF power perspective since refocusing pulses are not required. However, given that the signal is governed by the more rapid $T_2^*$ decay, it does suffer from reduced signal-to-noise ratio (SNR) relative to spin echo approaches. Fitting complex-valued data to the multi-exponential decay model has been shown to increase the robustness of MWF estimates made with this approach (Nam et al., 2015b).

Rather than modelling distinct tissue compartments solely from the decay of the transverse magnetisation, the mcDESPOT approach integrates spoiled gradient echo (SPGR) and balanced steady-state free precision (bSSFP) images, acquired with different nominal flip angles, to fit a two compartment model of the steady state signal (Deoni et al., 2008). The combination of these two acquisition types allows both $T_1$ (SPGR) and $T_2$ (bSSFP) to be estimated (Deoni et al., 2013). In the mcDESPOT model distinct relaxation times are determined for a fast and a slow relaxing pool, as well as the exchange rate (k), or residency time ($\tau$) of the two pools in the condition of chemical equilibrium. The fast relaxing pool is subsequently assumed to be myelin-associated water allowing the MWF to be quantified. The relaxation and exchange of these two pools is modelled using the Bloch-McConnell equations, which allows analytical solutions for the steady state signal to be derived (McConnell, 1958; Liu et al., 2016). Fitting the acquired data to these signal models requires seven distinct model parameters to be estimated: $T_1$, $T_2$ and fractional amplitude for each compartment as well as the exchange between them.

*2.2.2 BPF based on Magnetisation Transfer*

Like PD mapping, magnetisation transfer (MT) based approaches simplify the characterisation of white matter to two distinct pools (Fig. 1c). In this case one is comprised of an aqueous environment and the other a non-aqueous environment that, in the context of g-ratio mapping, is assumed to be associated with myelin. Unlike "free" water, such as found within the intra- or extra-cellular compartments, that has a sharp resonance linewidth, the "bound" non-aqueous protons have a much more heterogeneous microenvironment leading to a much broader range of MR frequencies and by consequence a very short $T_2$ in the range of tens of microseconds, meaning that the transverse magnetisation component is undetectable with MRI, unless ultra-short TE approaches are adopted (Sheth et al., 2016; Jang et al., 2020; Weiger et al., 2020). However, given its broad range of resonant frequencies, this bound pool can be selectively saturated through the application of an off-resonance radiofrequency pulse prior to conventional excitation and signal detection. This pre-pulse can selectively saturate the longitudinal magnetisation of the bound pool while leaving the free pool largely unaffected. Subsequently, the process of magnetisation transfer (MT), primarily occurring through dipolar coupling between the bound and free pools, leads to an observable reduction in the measured signal intensity (Wolff and Balaban, 1989; Sled and Pike, 2001; Sled, 2018; van Zijl et al., 2018). MT techniques capture the relative proportion of magnetisation in the bound pool relative to the free pool through the pool size ratio ($PSR = \frac{f_B}{f_W}$ (Sled and Pike, 2001) and Fig. 1c). Analogously to the MWF in MWI, the BPF, is defined as the magnetisation fraction within the bound pool relative to the total magnetisation in both pools (Sled, 2018): $BPF = \frac{f_B}{f_W + f_B}$ in Fig. 1c. In the first g-ratio mapping studies, the measured BPF was calibrated against histological data to convert it to an estimate of the MVF and combined with a diffusion-based measure of the FVF to estimate the g-ratio (Stikov et al., 2011, 2015).

The simplest means of probing the macromolecular bound pool via MT is to acquire an image using a pre-pulse with a single off-resonance frequency interleaved with a standard excitation pulse. The magnetisation transfer ratio (MTR) is defined as the normalised signal decrease relative to a reference image with only the standard excitation pulses (Henkelman et al., 2001). While this measure has been shown to be reflective of myelin content via histological analysis (Schmierer et al., 2004) hardware dependencies remain and reduce its comparability across individuals (Callaghan et al., 2015). A more robust measure that incorporates correction for both spatially varying $T_1$ and $B_1^+$ effects is the magnetisation transfer saturation (MTsat). This measure quantifies the percentage saturation per TR of the steady state SPGR signal that would result from a dual excitation sequence, and is dependent on the BPF (Helms et al., 2008). MTsat has been shown to be more robust to $B_1^+$ inhomogeneity than MTR (Callaghan et al., 2015) and to empirically correlate with the pool size ratio, F (Campbell et al., 2018).

More comprehensive modelling of the two magnetisation pools is obtained through quantitative MT (qMT) imaging. This approach aims to separate the contributions of the free and bound pools by explicitly modelling the distinct $T_1$ and $T_2$ relaxation times of the pools and incorporating the exchange between them, under the assumption of chemical equilibrium. The absorption lineshape of the bound pool must also be modelled, and is often assumed to be super-Lorentzian, with a $T_2$ in the region of tens of microseconds (Morrison and Henkelman, 1995). With this approach, the BPF can be estimated from the fractional magnetisation contributions of the two pools. To estimate this extended set of parameters, multiple images, sampling the so called z-spectrum, are acquired, each using a pre-pulse with a different off-resonance frequency (Sled and Pike, 2001; Cabana et al., 2015; Sled, 2018).

An intriguing, but not yet validated, approach that has also been used in the context of g-ratio mapping is to use multi-compartment Bloch simulations to model the myelin volume fraction within the voxel directly (Warntjes et al., 2016; Hagiwara et al., 2017).

*2.2.3 Protocol Considerations for MVF mapping*

A range of different protocols can be used to estimate the proton density, and by consequence the macromolecular tissue volume (Warntjes et al., 2007; Volz et al., 2012; Baudrexel et al., 2016; Mezer et al., 2016; Wang et al., 2018; Callaghan et al., 2019; Lorio et al., 2019). This approach requires an estimate of the receiver field sensitivity, $R$, which can be obtained by constrained model fitting or measurement (Mezer et al., 2016). The normalisation step to express PD as a fraction, or more commonly a percentage, of the concentration of protons in pure water requires a reference region to be defined, e.g. within the CSF-filled ventricles. However, the optimal choice of the normalisation region will depend on the acquisition scheme since sufficient SNR is required for robust estimation (CSF was used in (Berman et al., 2018) and white matter in (Ellerbrock and Mohammadi, 2018a)). The accuracy and precision of the PD estimation will in turn dictate the accuracy and precision of the MTV estimate. The mapping of PD was introduced in the context of fully relaxed signal (i.e. TR >> $T_1$). However, for reasonable scan times, this requirement can be relaxed, but in this case it is necessary to correct for spatially varying $T_1$ recovery.

Multi-compartment MWI necessitates short echo times to adequately sample the decay of the short $T_2$ myelin-associated water compartment (Whittall et al., 1999). This extends the minimum achievable TR and can lead to long acquisition times, particularly for spin echo based approaches, unless spatial coverage or resolution are sacrificed. Acquiring multiple spin echoes in a single readout increases temporal efficiency, but the train of pulses can lead to the refocusing of echoes from unwanted pathways, i.e. the production of stimulated echoes, when the transmit field, $B_1^+$, is inhomogeneous. Correction schemes based on simulating the impact of these echoes (e.g. (Lebel and Wilman, 2010)) have been proposed and can be incorporated into the fitting procedure. 2D slice-selective approaches are also vulnerable to distorted slice profile effects, which can be mitigated either by modifying the sequence to ensure a sufficiently broad refocusing width, or by accounting for the effect during processing (Lebel and Wilman, 2010; Nöth et al., 2017). The large number of refocusing pulses also increases the specific absorption rate (SAR) of the sequence and can be a limiting factor at higher field strengths.

MTR and MTsat are time efficient means of quantifying the effect of magnetisation transfer. As highlighted earlier, MTsat is more hardware robust. In addition, high resolution maps can be obtained with whole brain coverage in reasonable scan times making it particularly appealing for clinical studies. This efficient method was used in the first group study mapping the g-ratio *in vivo* (Mohammadi et al., 2015). However, a limitation of these rapid approaches is that they are semi-quantitative. The saturation of the bound pool, and therefore of the free pool via magnetisation transfer, will depend on the particular off-resonance pulse used, most notably the power and offset frequency. For further details, acquisition protocols and software for estimating this parameter see e.g. (Tabelow et al., 2019).

qMT approaches circumvent this limitation by quantifying specific physical parameters. However, the extended datasets required to fit the full qMT model fitting lead to a trade off between scanning durations and spatial resolution and/or coverage. To constrain the model fits, parameters can be fixed, e.g. the T1 of the free and bound pool can be set equal to each other, or an "observed" $T_1$ can be separately measured and integrated into the fitting to relate the $T_1$ times of the bound and free pools. For further details and software available for fitting such models, see e.g. (Cabana et al., 2015).

Clearly brain tissue can be characterised by a very broad range of physical parameters. The multi-parameter mapping (MPM) quantitative MRI protocol offers a comprehensive approach providing high resolution, whole brain estimates of (single compartment) $T_1$, $T_2^*$, PD, MTV and MTsat, with correction for transmit and receive field effects, in clinically feasible scan times (Weiskopf et al., 2013; Callaghan et al., 2019; Tabelow et al., 2019). As such it provides simple proxies for both the macromolecular (via MTsat & MTV) and free water pools (PD) in a single protocol.

## 2.3  Axonal Volume Fraction and Fibre Volume Fraction

Diffusion MRI is the method of choice to separate the intra- and extra-axonal tissue compartments ($f_{AW}$ and $f_{EW}$, Fig. 1d) because of the distinct diffusion properties of water in these compartments. However, as detailed above, the myelin-associated water compartment has a short $T_2$. This means that diffusion-weighted MRI is insensitive to myelin water because of the comparatively long minimum echo time required to accommodate the application of diffusion gradients. Nonetheless, there are several different diffusion-based approaches available to probe the intra-axonal tissue compartment. Detailing each of these goes beyond the scope of g-ratio mapping. For those interested in more details, we refer to other excellent reviews (e.g. (Alexander et al., 2019; Novikov et al., 2019)). Here we will specifically focus on those approaches that have been used to date to estimate the intra-axonal volume fraction for the purpose of computing the aggregated g-ratio. These studies can be subdivided into two categories: the studies that have used standard DTI data and those that have used multi-shell (and even more advanced) diffusion MRI protocols. Each category will be discussed in turn.

### 2.3.1 FVF from DTI data

The first category of studies requires only a limited set of measurement parameters, including only a single b-value and a modest number of diffusion directions, as defined by the DTI protocol. Therefore, these studies refrain from explicitly modelling more than one tissue compartment. A feature of these studies was the interpretation of diffusion-MRI based measurements of the axonal compartment as the FVF rather than the AVF, which, given the insensitivity of diffusion MRI to myelin water, is probably incorrect as we will discuss further in the next section.

*DTI:* The first g-ratio mapping study by Stikov et al. (Stikov et al., 2011) used simulations, in which axons were modelled as straight, parallel cylinders to establish a second order relationship between the fractional anisotropy (FA) of the diffusion tensor and the total $FVF$. The assumption of straight and parallel cylinders, however, restricted the application of this model to white matter regions with well aligned fibres. As a result, it has only been applied to the corpus callosum to date (Stikov et al., 2011; Berman et al., 2018).

*TFD:* The diffusion model used by Mohammadi et al. (Mohammadi et al., 2015) for g-ratio mapping was based on the tract-fibre density (TFD), which is not restricted to well-aligned fibre pathways and thus could be applied across the whole brain. The TFD was derived from fibre orientation distributions (Reisert et al., 2013) and assumed to be directly proportional to the $FVF$. The proportionality constant that related TFD to $FVF$ was combined with the calibration coefficient that related the MTsat myelin marker used to capture $MVF$. The resulting calibration constant was estimated by referencing against a ground truth g-ratio value from literature (Mohammadi et al., 2015). This calibration approach will be further discussed in the context of myelin biomarkers in section 3.2. However, Ellerbrock et al. (Ellerbrock and Mohammadi, 2018a) recently showed the TFD-based $FVF$ parameter to be less stable in terms of repeatability and comparability than $FVF$ estimates derived from the Neurite and Orientation Dispersion in Diffusion Imaging (aka NODDI) model (Zhang et al., 2012), discussed in more detail in the next section.

*2.3.2  AVF from multi-shell diffusion MRI data*

Using a more extensive set of experimental measurement parameters, i.e. multiple b-values  (aka diffusion shells), allows the second category of studies to use a more principled model for the diffusion signal, the so-called "standard model" (Novikov et al., 2019).  The standard model is built upon well-established signal models for two tissue compartments (for a summary see, e.g., (Novikov et al., 2019)), the axonal ($A$) and extra-cellular ($E$) compartments (Fig. 3a.ii). A restricted signal component is assumed to come from the axonal compartment, which is modelled as impermeable sticks (Fig. 3a.iii). A hindered signal component describes the extra-cellular space, which is modelled using a 3D anisotropic diffusion tensor. Such an example is depicted in Figure 3c for the White

Matter Tissue Integrity (WMTI) model, showing the axially-symmetric ellipsoidal tensor composed of axial ($D_{E,||}$) and perpendicular ($D_{E,\perp}$) extra-cellular diffusivities.

In contrast to g-ratio studies based on DTI data, those using multi-shell diffusion MRI data also acknowledge the fact that the direct contribution of myelin water in the diffusion MRI signal is negligible (Fig. 3a.ii). As a consequence, their models take into account that the axonal compartment estimated from the visible MRI signal in a typical diffusion experiment is not $AVF = \frac{AVF}{MVF+AVF+EVF}$ (Fig. 3a.i) but rather the axon water fraction ($AWF$), i.e. intra-axonal signal divided by the signal from the extra- and intra-cellular space: $AWF = \frac{f_{AW}}{f_{AW}+f_{EW}}$ (Fig. 1d) and thus $AWF = \frac{AVF}{AVF+EVF}$ (Fig. 3a.ii). These studies follow the suggestion of Stikov et al. (Stikov et al., 2015) to estimate $AVF$ by rescaling the $AWF$ accounting for the unsampled $MVF$, i.e.:

(2) $AVF = (1 - MVF)\, AWF$.

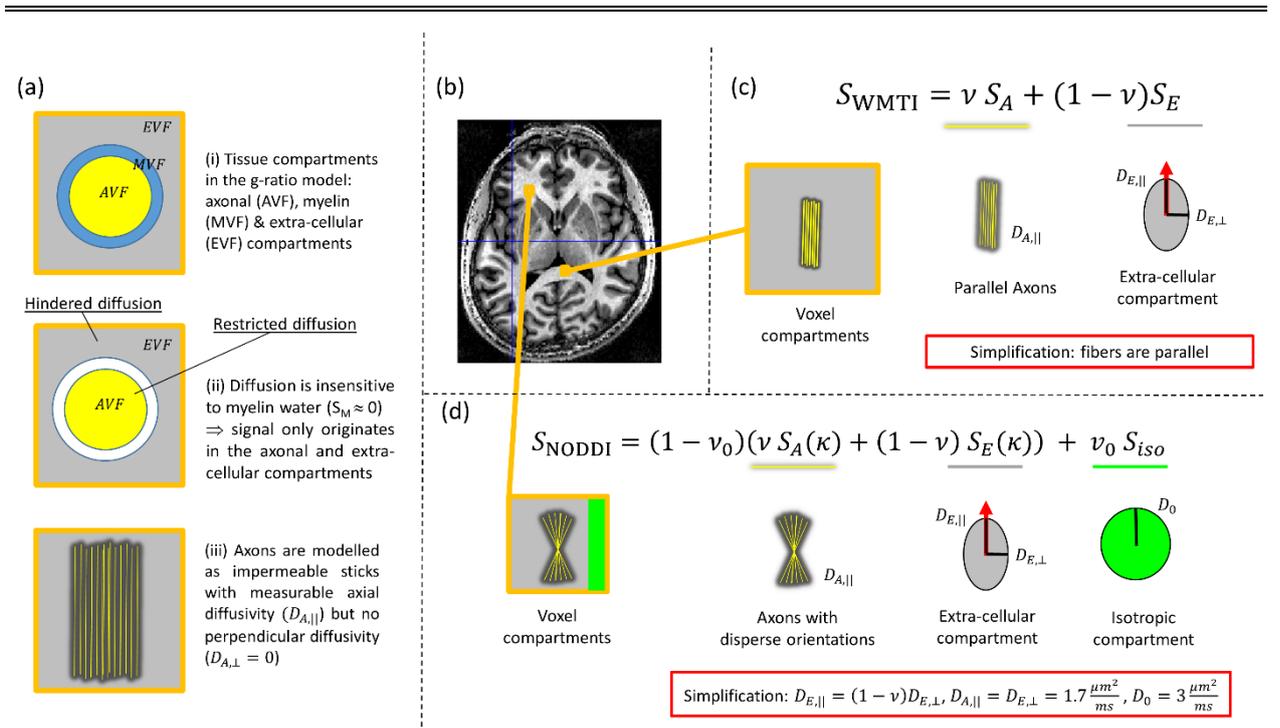

**Figure 3**: Depicted are the compartments of the g-ratio white matter (WM) tissue model as seen by diffusion MRI (a). An axial view of the human brain (b) is used to indicate WM regions where two example signal models that have been used to estimate the axonal water fraction ($AWF$) are applicable (c,d). (a.i): The cross-section of a representative myelinated axon and the associated tissue compartments in the g-ratio model: axonal ($AVF$ in yellow), myelin ($MVF$ in blue), and extra-cellular ($E$ in gray) volume fractions ($VF$). (a.ii): Only two out of three compartments of (a.i) contribute to the diffusion signal: $S_E$ and $S_A$. The contribution from myelin is negligible because of its short T$_2$, i.e. $S_M = 0$. (a.iii): Typical diffusion models assume that the axonal compartment is composed of a population of sticks (depicted lengthwise in yellow) in which there is measureable diffusivity only along the length of the sticks (i.e. $D_{A,\|} > 0$ and $D_{A,\perp} \approx 0$). (c): The White Matter Tissue Integrity (WMTI) model is comprised of distinct signal contributions from within axons ($S_A$) and from the extra-cellular space ($S_E$) with corresponding signal fractions: $v$ from $S_A$ and $(1 - v)$ from $S_E$ and compartment-specific diffusivities: $D_{A,\|}$, $D_{E,\|}$ and $D_{E,\perp}$. However, WMTI can only be applied to WM regions with well-aligned fibre pathways because it assumes parallel sticks thereby excluding disperse orientations. To satisfy this model assumption it has been used only within the corpus callosum (see (b)). (d) The Neurite and Orientation Dispersion in Diffusion Imaging (NODDI) model is comprised of axonal ($S_A$) extracellular ($S_E$) and isotropic ($S_{iso}$) signal compartments and estimates signal fractions for each (i.e. $v$, $(1 - v)$ and $v_0$ respectively). To improve fitting stability, the NODDI model makes very strong assumptions, e.g.: the intra- ($D_{A,\|}$) and extra-axonal parallel diffusivities ($D_{E,\|}$) are assumed to be the same and are fixed, as is the diffusivity of the isotropic compartment ($D_0$) corresponding to CSF. The parallel and perpendicular diffusivities are assumed to be related via the tortuosity model: $D_{E,\perp} = D_{E,\|}(1 - v)$. The depicted values are for the *in vivo* case. However, NODDI does not assume parallel fibres, but rather accounts for fibre dispersion ($\kappa$), which is described by a Watson distribution (Stoyan, 1988). NODDI can therefore be used in regions with more disperse fibre orientations (as depicted in (b)).

*WMTI:* The WMTI model (Fieremans et al., 2011) contains signal contributions from intra-axonal ($S_A$) and extra-cellular ($S_E$) compartments and is therefore directly related to the "standard model" of diffusion MRI. In this model, the signal fraction of sticks ($v = \frac{f_{AW}}{f_{AW}+f_{EW}}$, Fig. 1d) is directly used as proxy for $AWF$ while $1 - v$ ($= \frac{f_{EW}}{f_{AW}+f_{EW}}$, Fig. 1d) estimates the extra-cellular water fraction (Fig. 3c). In addition to $AWF$, WMTI simultaneously estimates the intra-axonal diffusivity ($D_{A,||}$) and two extra-cellular diffusivities ($D_{E,\perp}$ and $D_{E,||}$) of an axially-symmetric ellipsoidal tensor. However, it assumes parallel fibres and therefore has applied been only to the corpus callosum (West et al., 2018a).

*mcSMT*: Like WMTI, the multi-compartment Spherical Mean Technique (mcSMT) model developed by Kaden et al. (Kaden et al., 2016) is based on the "standard model". But, instead of assuming parallel fibres, it uses the SMT to factor out the contribution of fibre orientation. As a result, it can be applied to the whole brain. Similar to the WMTI model, mcSMT estimates the signal fraction of the intra-axonal space, $v$. This has been used as a proxy for the $AWF$ in g-ratio mapping (West et al., 2018a). In the mcSMT model, the intra- and extra-cellular parallel diffusivities are assumed to be equal ($D_{A,||} = D_{E,||}$) and the tortuosity model (Szafer et al., 1995) is used to relate the extra-cellular parallel and perpendicular diffusivities to each other via $v$: $D_{E,\perp} = (1-v)D_{E,||}$.

*NODDI:* The most commonly used method to estimate the $AWF$ in g-ratio mapping has been the NODDI model (Fig. 3d, (Zhang et al., 2012; Stikov et al., 2015)). As compared to the aforementioned models, NODDI is a 3-compartment signal model. It not only models the two signal compartments from the intra-axonal and extra-cellular space but also a third isotropic signal component ($S_{iso}$ with an associated signal fraction $v_0 = \frac{f_{CSF}}{f_{AW}+f_{EW}+f_{CSF}}$, Fig. 1d) to account for any partial-volume contamination by freely diffusing water, e.g., as in CSF. To compensate for the increased number of model parameters and stabilize model fitting, the diffusion constants are fixed (Fig. 3b). To this end, as in the mcSMT model, the tortuosity model is used to relate the extra-axonal diffusivities via $v$: $(D_{E,||} = (1-v)D_{E,\perp})$. Moreover, the intra-axonal and extra-axonal parallel diffusivities are assumed to be equal ($D_{A,||} = D_{E,||}$) and have a predefined value, as does the diffusivity of the isotropic compartment ($D_0$). To account for the 3-compartment nature of NODDI, Stikov et al. (2015) suggested the following relation between the NODDI signal fractions (Fig. 3d) and the $AWF$: $AWF = v(1 - v_0)$. By scaling $v$ with $1 - v_0$, ($v(1 - v_0) = \frac{f_{AW}}{f_{AW}+f_{EW}}\left(1 - \frac{f_{CSF}}{f_{AW}+f_{EW}+f_{CSF}}\right) = \frac{f_{AW}}{f_{AW}+f_{EW}+f_{CSF}}$) such that the intra-axonal signal fraction is corrected for the contribution of the CSF compartment, to ensure the g-ratio WM model assumption, i.e. $AVF + MVF + EVF = 1$ (Fig. 1d).

NODDI accounts for fibre dispersion using the single-parameter Watson distribution (Stoyan, 1988; Jespersen et al., 2012), making it applicable for whole brain g-ratio mapping.

*CHARMED:* Compared to other diffusion models that have been used for g-ratio mapping, the Combined Hindered and Restricted Models of water diffusion (CHARMED) approach makes the fewest assumptions. It models diffusion in the extra-cellular space by a full ellipsoidal tensor (whereas the NODDI and WMTI models assume an axially-symmetric ellipsoid), and, in principle, it can account for crossing fibre configurations (Assaf et al., 2004; Assaf and Basser, 2005) unlike the standard NODDI approach. The CHARMED model can be further extended to additionally estimate axon diameters (e.g. (Assaf et al., 2008; Alexander et al., 2010; Huang et al., 2016)). This has been used by Duval et al. for g-ratio mapping in the spinal cord (Duval et al., 2017) and by Yu et al. (Yu et al., 2019) in patients with multiple sclerosis. However, such a protocol requires more extensive (and time-consuming) data acquisition.

*2.3.3 Protocols for AVF mapping*

While the first category of studies requires only a standard single-shell DTI protocol (Stikov et al., 2011; Mohammadi et al., 2015; Berman et al., 2018), the minimum requirement protocol for the second category of studies depends on the model to be used for AWF mapping. The WMTI model parameters can be estimated from the diffusion kurtosis tensor measurement (Fieremans et al., 2011; Jespersen et al., 2018). The NODDI, mcSMT, and WMTI model parameters can be estimated from a two-shell diffusion MRI protocol composed of a "lower" ($b \sim 1 \frac{ms}{\mu m^2}$) and a "higher" diffusion weighting ($b \sim 2 \frac{ms}{\mu m^2}$) [1]. In contrast to the aforementioned models, the CHARMED model typically requires a more extended diffusion MRI protocol: Drakesmith et al. used a five shell diffusion MRI dataset for g-ratio weighted imaging (Drakesmith et al., 2019). Extending the CHARMED model to also estimate axon diameters requires an even more advanced protocol where the b-values and additional diffusion parameters such as diffusion sensitization times also have to be changed ( see (Duval et al., 2017) for g-ratio mapping).

Typical protocol-associated issues that can introduce biases are: ceiling effects (i.e. $v = 1$ in white matter, which can be encountered with NODDI if b-shells are sub-optimally sampled (recommendations for optimal sampling provided in (Zhang et al., 2012)). Rician noise in low SNR data can also lead to bias in AWF estimation can propagate into the AWF parameters from WMTI. Mapping accurate AWF parameters in the spinal cord comes with additional challenges because of increased susceptibility to nonlinear motion (e.g. due to swallowing, (Yiannakas et al., 2012)), physiological noise (e.g. (David et al., 2017)), or partial volume effects due to its small size (1 cm in diameter).

---

[1] Note that these parameters are for in vivo imaging and will be different for ex vivo MRI. For example, in the study by (West et al., 2018a) the low and higher diffusion weighting were at $b \sim 3 \frac{ms}{\mu m^2}$ and $b \sim 6 \frac{ms}{\mu m^2}$, respectively.

## 3   Challenges for aggregated g-ratio mapping

The most important prerequisite of g-ratio mapping with MRI is that the biomarkers of MVF and AVF be accurate. Two key requirements for an accurate biomarker are model validity and a one-to-one correspondence between the MRI-biomarker and the gold standard MVF and AVF. While the first point can be investigated by theoretical evaluation of the model, the second point is typically not fulfilled necessitating a calibration step. Another important challenge is related to imaging artefacts and their impact on the multi-modal combination of MVF and AVF biomarkers. In this section, we will first discuss the question of model validity associated with MRI-based MVF and AVF biomarkers, then we will use a simulation experiment based on *ex vivo* data to improve our understanding of the calibration step, and finally we discuss imaging artefacts associated with the multi-modal combination of MRI data.

### 3.1   Model Validity

It is important to bear in mind that "all models are wrong but some are useful" [2]. In the following sections we will cover some of the key model assumptions made to facilitate *in*

---

[2] The aphorism is generally attributed to the statistician George Box, although the underlying concept predates Box's writings (https://en.wikipedia.org/wiki/All_models_are_wrong).

*vivo* mapping of the AVF and MVF and enable g-ratio mapping. We will also discuss the consequent limitations of application.

*3.1.1 MVF models*

The simplest model for estimating $f_B$ is based on PD mapping, in which a mono-exponential, i.e. single water compartment, is typically assumed when extrapolating the signal to a TE of 0ms to remove confounding $T_2^{(*)}$ decay. This is clearly not valid and constituent water compartments within a voxel will have variable influence depending on the echo times and spacings used (Whittall et al., 1999). This will be the case for both PD mapping and MWI. In general, longer apparent $T_2^{(*)}$, and smaller fractional contribution from short $T_2$ components, are observed as the first TE is increased (Cercignani et al., 2018). It is also important to fully sample the decay, which requires sufficiently long echo times to capture any slowly decaying compartments. However, when fitting magnitude data with long echo times, significant biases can be introduced by the Rician noise distribution and greatly alter the measured $T_2^{(*)}$ values (Bjarnason et al., 2013). As noted earlier, complex-valued fitting can be particularly beneficial when the MWF is characterised via the shorter T2* decay (Nam et al., 2015b) of gradient echo imaging. Moreover, it has recently been shown that MWF depends on the orientation of fibres with respect to the external magnetic field (Birkl et al., 2020). Sensitivity to $B_0$ inhomogeneity can also bias model fits as can phase errors caused by physiological effects, such as breathing and eddy currents (Nam et al., 2015a) and motion, which distorts the decay (Magerkurth et al., 2011). Vulnerability to physiology and motion, together with partial volume effects, are particularly problematic for spinal cord imaging (Duval et al., 2017, 2018; Hori et al., 2018). More generally, these potential sources of artefact can manifest differently *in vivo* and *ex vivo*, meaning that while some techniques may work well in post mortem data, e.g. achieving cross-validation with histological data, they may not necessarily work well *in vivo*.

Models assuming two pools, either distinct non-exchanging water pools in myelin water imaging (Fig. 1b) or a bound and free pool that interact via magnetisation transfer (Fig. 1c) are also limited by the fact that they do not describe the full complexity of the tissue's microstructure. Higher numbers of pools, are undoubtedly present (c.f. even the simplified model of Fig. 1a) but are unlikely to be distinguishable based on distinctly observable relaxation behaviour either because of exchange or because it would require unattainable measurement precision. Simulation studies of more complete models have helped us to better understand the limitations of these simplifications.

In MWI, a slow exchange rate, is central to the possibility of differentiating water pools, and their fractional sizes, based on experimentally distinguishable $T_2$ times. As the exchange rate increases to a more intermediate regime, distinct compartments may still be discernible, but the relaxation times will appear reduced, as will the MWF (Does, 2018). The situation is further complicated by the presence of noise, which, even at low levels, can further broaden the distribution of apparent relaxation times, and even lead to distinct water environments merging in the three pool case (Does, 2018). The rate of magnetisation transfer exchange between macromolecular and water pools is an order of magnitude larger than the diffusion-driven exchange rate between water compartments (c.f. non-directional exchange rates of $10s^{-1}$ and $100s^{-1}$ respectively, (Levesque and Pike, 2009)). Theoretical analysis of a four pool model (analogous to Fig. 1a) has also shown that inter-compartmental exchange could substantially alter the estimated MWF, but that the qMT-based BPF is more robust (Levesque and Pike, 2009). In support of this theoretical analysis, much greater variation in MWF than BPF has been seen in the spinal cord, not only *ex vivo* (Dula et al., 2010) but also *in vivo* (Harkins et al., 2012). The variability observed across tracts was consistent with variable exchange due to differences in axon diameter and myelin thickness, the key determinants of the g-ratio. Much of the extensive validation work for the MWI technique has been conducted *ex vivo*, and often with samples at room temperatures. Both of these factors serve to slow the rate of exchange increasing the validity of the slow exchange assumption (Does, 2018). Therefore, one must again exercise caution extrapolating the validity of MWF metrics from *ex vivo* findings to the *in vivo* situation.

Although these three and four pool models are likely to be closer to the true tissue microarchitecture, inversion of such a complex model would be difficult in terms of both precision and bias. Indeed, even in the context of the two pool models that have been used to date for g-ratio mapping, the parameterisation must be supported by the data. The comparatively high parameterisation of the mcDESPOT model has necessitated the use of advanced fitting procedures, such as stochastic genetic or region contraction algorithms (Deoni et al., 2008, 2013). The achievable precision and accuracy of the approach has been called into question (Lankford and Does, 2013; West et al., 2019) and it has been shown to suffer from degeneracy when seeking to determine optimal model parameters, which is only resolved by using a simpler model, excluding exchange (West et al., 2019). A common requirement of all model types, including those capturing the AWF, is that any fixed parameters, e.g. as might be assumed in qMT models where the $T_1$ of the free and bound pools may be assumed to be equal (Cabana et al., 2015), be appropriate to the population under consideration be they adults, children or indeed patients.

While it is also incorrect to assume that the non-aqueous compartment of tissue is entirely comprised of myelin, this has been shown to be the dominant source of the MT contrast mechanism in WM (Eng et al., 1991). In reality, the bound pool can be associated not only with the lipids and proteins of the myelin sheath, but also with any other macromolecule-bound protons (see Fig. 1a), e.g. glial cells (MacKay and Laule, 2016).

MWF will not only capture water within myelin sheaths surrounding axons but also that associated with any myelin debris in pathological cases, as has been shown in peripheral nerve (Webb et al., 2003). Similarly, MT-based measures lack specificity. Hence it should be borne in mind that although alterations in myelin content will change the measured MT effect, an alteration in MT effects cannot be uniquely attributed to a change in myelin and may be driven by other macromolecular changes. The derived MVF is also used to correct for the fact that the diffusion signal is insensitive to this compartment (by rescaling AWF). However, this neglects the non-myelin-macromolecular contribution within the imaging voxel, i.e. $f_{BNM}$ (Fig. 1a).

Finally, MWF is most commonly estimated within the myelin-rich white matter but has been shown to be significantly lower in grey matter (MacKay et al., 2006; MacKay and Laule, 2016). In this case, a further difficulty relates to whether or not sufficient sensitivity can be achieved *in vivo* to reliably estimate MWF in grey matter.

### 3.1.2 AWF models

Examples of strong simplifications used by the AWF models are that: the restricted compartment is solely associated with axons that can be modelled as impermeable sticks without cross-section, and that diffusion in the extra-cellular space is assumed to be Gaussian. The assumption that the restricted compartment is solely associated with axons is expected to be approximately correct in white matter because the density of other cells is small relative to the density of axons. In grey matter, however, the restricted diffusion signal will depend not only on the axonal compartment but on the density of cells of all types (soma and glia) as well. Thus, the simple three compartment approach of non-CSF tissue stipulating that AVF+MVF+EVF=1, where the combination of one diffusion and one myelin biomarker can be used to estimate AVF, no longer holds because the diffusion biomarker derived from the restricted signal component will be weighted by both AVF and EVF.

In addition to these model limitations, there is another problem associated with all of the approaches used for g-ratio mapping to date: they are based on the standard model comprised of compartments accounting for restricted and hindered diffusion. This model is known to suffer from a degeneracy of parameter estimates (Jelescu et al., 2016a) when measured with a linear diffusion weighting approach, i.e. the typical Stejskal and Tanner (Stejskal and Tanner, 1965) diffusion weighting scheme, which has been the case for all the aforementioned g-ratio mapping studies.

Alternatively, prior assumptions motivated by the biological composition of the tissue can be imposed to stabilize the parameter estimation. The NODDI, mcSMT, and WMTI models make particularly strong use of prior assumptions to allow the remaining model parameters to be estimated from data that can be acquired in a clinically feasible imaging time (see section 2.3.3). NODDI and mcSMT use a tortuosity model (Szafer et al., 1995) to relate the perpendicular extra-axonal diffusivity to the parallel extra-axonal diffusivity scaled by "one minus the neurite density": $(D_{E,\perp} = (1-\nu)D_{E,||})$, i.e. the higher the neurite density in the tissue the lower the perpendicular diffusivity. As discussed in (Jelescu et al., 2015), the tortuosity model, however, should relate extra-axonal diffusivities to the extra-cellular space $(EVF = 1 - FVF)$ rather than signal fraction of the hindered compartment, $\nu$. In other words, the relationship between the parallel and perpendicular extra-cellular diffusivities should effectively be: $D_{E,\perp} = (1-FVF)D_{E,||}$. Moreover, NODDI and mcSMT impose a one-to-one scaling between the intra- and extra-cellular parallel diffusivities: $D_{A,\parallel} = D_{E,\parallel}$. While mcSMT estimates the remaining diffusivity, NODDI fixes it to a constant value (for *in vivo* healthy adults the diffusivities are usually assumed to be: $D_{A,\parallel} = D_{E,\parallel} = \frac{1.7 \mu m^2}{ms}$ and $D_0 = \frac{3 \mu m^2}{ms}$). This latter assumption would be problematic if different populations were studied, for which these diffusivities were not applicable, e.g. children, patients, or postmortem brains. WMTI, on the other hand, assumes that all fibres are aligned in parallel restricting its application to anatomical regions that support this assumption, e.g., it has been applied to the corpus callosum (West et al., 2018a). But, even in the corpus callosum axons are not necessarily aligned fully in parallel. This might be another reason (in addition to fixed diffusivities used in NODDI) for the systematically smaller AWF estimates when using WMTI as compared to NODDI as reported, e.g., in (Jelescu et al., 2015). Of course, this list of model assumptions that should be borne in mind is not exhaustive (Jelescu and Budde, 2017; Novikov et al., 2019). The Watson distribution used in NODDI can model fibre dispersion in a single fibre population, but cannot describe more complex fibre scenarios, such as crossing fibres. Nevertheless, it accounts, to a certain degree, for the variability of fibre-alignment within fibre pathways and thus might be better suited for whole brain g-ratio mapping than models that assume strictly parallel fibre configurations.

### 3.2  Calibration for MVF

Assuming that the diffusion-based AWF is accurate[3], the relation between the myelin biomarker and the MVF still needs to be established via a calibration step. This calibration is particularly important since it is not only required to quantify the MVF, but also to convert the AWF to AVF (Eq. 2). Histological investigations suggest that the relationship between typical myelin biomarkers (which we will collectively denote $M_{MRI}$ in this section) and the MVF is linear (Fig. 4, (West et al., 2018b)):

(3) $MVF_{MRI} = \alpha M_{MRI} + \beta$,

where $\alpha$ and $\beta$ are unknown coefficients that need to be calibrated. It is expected that these coefficients will depend on instrumental variables and may therefore vary with MR systems, sequence parameters, as well as myelin biomarker models. Such dependency clearly limits the reproducibility and comparability of the MR-based g-ratio. Using simulations, Campbell et al. (2018) demonstrated that imperfect calibration can not only introduce a bias in the g-ratio, but can even cause the g-ratio to depend on the fibre volume fraction, negating the major strength of the g-ratio, i.e. that it is independent of FVF. Their simulations revealed that this dependence was different if the miscalibration was present only in the offset or only in the slope and coined the phrase *aggregated g-ratio weighted imaging* (Campbell et al., 2018).

---

[3] This assumption is probably wrong – see model validity in section 3.1.

To reduce these dependencies, two calibration methods have been used for *in vivo* g-ratio mapping. These have utilised a region of interest (ROI) in which either (a) the myelin biomarker was calibrated against the reference $MVF$, first employed by (Stikov et al., 2015) or (b) the measured g-ratio was calibrated against the reference g-ratio, first employed by (Mohammadi et al., 2015). We refer to these approaches collectively as **single-point calibration** methods since both are calibrating against a single reference value. Reformulating Eq. 3 within a specific ROI, it is clear that the single-point calibration methods estimate one effective proportionality constant ($\alpha_{eff}$), i.e.:

(4) $MVF_{MRI}(ROI, \alpha_{eff}) = \left(\alpha + \frac{\beta}{M_{MRI}(ROI)}\right) M_{MRI}(ROI) \equiv \alpha_{eff}(\alpha, \beta, M_{MRI}) M_{MRI}(ROI)$.

From Equation (4) it is clear that the single-point calibration methods are insufficient to establish a one-to-one correspondence between the MVF and the MRI-based myelin biomarker. One problem, for example, could be that $\alpha_{eff}$ will depend on the myelin biomarker within the reference ROI if $\beta \neq 0$ (see Eq. (4)). The MVF-based single-point calibration method would simply set Equation 4 to a reference MVF value within the ROI: $MVF_{MRI}(ROI, \alpha_{eff,opt}) = MVF_{REF}$ and rearrange the equation with respect to $\alpha_{eff,opt}$. The g-ratio based single-point calibration would minimize the following equation:

$$(5) \quad \alpha_{eff,opt} = \min_{\alpha_{eff}} \| g(\alpha_{eff}, ROI) - g_{REF}(ROI) \|$$

with $g(\alpha_{eff}, ROI) = \sqrt{1 - \frac{MVF_{MRI}(ROI, \alpha_{eff})}{MVF(ROI, \alpha_{eff}) + (1 - MVF(ROI, \alpha_{eff})) AWF(ROI)}}$ where $g_{REF}(ROI)$ and $AWF(ROI)$ are the reference g-ratio and the measured AWF values within the ROI respectively.

The key questions that ensue from this single-point calibration are: what are the typical magnitudes of the slope $\alpha$ and offset $\beta$ in experimental conditions and therefore what is the magnitude of the error propagated by $\alpha_{eff,opt}$? How much does the MR-based g-ratio deviate from the ground truth? How large is this deviation relative to the expected dynamic range of the g-ratio, e.g. pathology-related differences?

Although the simulations in (Campbell et al., 2018) improved our understanding of the pitfalls of g-ratio mapping, they did not directly answer these questions. However, experimental data from the Does lab (Kelm et al., 2016; West et al., 2018b, 2018a) could help answer these questions. In those experiments, the authors reported the changes of the g-ratio and the associated myelin-volume fractions in a range of mouse models spanning hypo- to hyper-myelination using both MRI and electron microscopy. The MRI based data included three biomarkers of myelin content: MWF ($f_{MWF}$), BPF, and MTV.

Since in this case MTV was derived from the MWI experiment (i.e. with a multi-compartment model) we denote it $MTV_{MWI}$. In the following, we will use the range of reported values for MVF from histology, and AWF from diffusion MRI, to generate ground truth parameters for a subsequent simulation-based experiment.

We will evaluate the bias and error between the ground truth g-ratio, $g_{GT}$, and that obtained by calibrated MRI measures, $g_{MRI}$, using Bland-Altman analyses (Bland and Altman, 1986). The Bland-Altman plots (Fig. 5) depict the difference ($\delta_g = g_{GT} - g_{MRI}$) between the g-ratios as a function of their mean ($m_g = (g_{GT} + g_{MRI})/2$). The solid line indicates the mean difference $\langle \delta_g \rangle$, while the dashed lines indicate $\langle \delta_g \rangle$ plus/minus 1.96 times the standard deviation of the differences $\langle std_{\delta_g} \rangle$ to encompass 95% of the normal distribution. According to their original publication (Bland and Altman, 1986), $\langle \delta_g \rangle$ and $+/-1.96 \langle std_{\delta_g} \rangle$ can respectively be interpreted as the bias and error that would result if replacing $g_{GT}$ with $g_{MRI}$. **Bias** captures the offset from the ground truth g-ratio value, whereas **error** captures the deviation from a one-to-one relationship between the ground truth and the MR g-ratio. While a potential bias can be retrospectively corrected, the error in the g-ratio mapping method will define its sensitivity and ability to detect change or differences between individuals, groups or over time. Any error must be lower than the expected difference between groups or due to pathology if the g-ratio mapping method using MRI is to be of use to reliably assess these differences.

### 3.2.1 Ex vivo simulation experiment

To generate a realistic range of ground truth values for the MVF and AWF, we used the histology-based MVF values reported in (West et al., 2018b), which range from approximately 0.015 to 0.285 (cf. Fig. 7 in (West et al., 2018b)) and the MRI-based AWF values as reported in (Kelm et al., 2016), ranging from approximately 0.3 to 0.7 (cf. Fig. 13 in (Kelm et al., 2016)). Note all measurements are derived from the same animals. Then, we used Eqs. (1) and (2) to generate the ground truth g-ratio values ($g_{GT}$, which ranged from 0.79 to 0.97) [4]. To generate the MRI-based myelin marker, we used the linear relationships reported in (West et al., 2018b) between the histological MVF (here: the ground truth MVF) and three myelin biomarkers: $BPF = 0.45\ MVF + 0.086$ (Fig. 4a [5]), $f_{MWF} = 0.89\ MVF - 0.016$ (Fig. 4b), and $MTV_{MWI} = 0.75\ MVF - 0.047$ (Fig. 4c [6]). Note that the calibration of $MWF$ was independent of the experimental data in (West et al., 2018b) but based on literature values from an independent experiment. Therefore, $f_{MWF}$ was used in the following simulations instead of $MWF$. This was opposite to the calibrated $BPF$, which was estimated using the experimental data in (West et al., 2018b). Also note that the $MTV_{MWI}$ requires an intrinsic calibration to normalize the water content (see section 2.2.1).

In this simulation experiment, we compared $g_{GT}$ with the non-calibrated g-ratio values ($g_{MRI}^{none}$) and with the calibrated g-ratio values ($g_{MRI}^{SPC}$) using either the g-ratio (Fig. 5b,e) or MVF (Fig. 5c,f) single-point calibration (SPC) methods (depicted as scatter in Fig. 5a-c and Bland-Altman plots in Fig. 5d-f). The SPC-reference values were based on the average in control mice (blue symbols in Fig. 7 (West et al., 2018b): $MVF_{REF} \approx 0.175$ and $g_{REF} \approx 0.85$). The index $MRI$ described the myelin biomarker that was used to generate the g-ratio: i.e. $g_{BPF}$ used $BPF$ (blue crosses), $g_{f_{MWF}}$ used $f_{MWF}$ (green crosses), and $g_{MTV_{MWF}}$ used $MTV_{MWI}$ (black crosses). When calculating the g-ratios, an upper and lower limit was applied meaning that if $g^2_{MRI} > 1$, the g-ratio value was set to one (because $\left(1 - \frac{MVF}{FVF}\right) \leq 1$) and if $g^2_{MRI} < 0$, the g-ratio was set to zero (because $\left(1 - \frac{MVF}{FVF}\right) \geq 0$). In the results, we report the bias and error of the Bland-Altman analyses relative to the dynamic range of simulated ground truth g-ratios: $g_{dyn} = \max(g_{GT}) - \min(g_{GT}) = 0.18$.

---

[4] Note that these values are larger than the g-ratios assessed via ex vivo histology in (West et al., 2018a), ranging from 0.75 to 0.88. This may be because the latter was determined from myelinated axons only. The g-ratios in this simulations, however, are based on the combination of histological MVF and diffusion MRI-based AWF. The diffusion-based AWF should be sensitive to both myelinated and unmyelinated axons (Beaulieu and Allen, 1994a; Beaulieu, 2002). Since the g-ratio of unmyelinated axons is 1, a larger ground truth g-ratio is expected than the EM-based values in (West et al., 2018a).

[5] Note that the linear equation reported in Figure 7 (West et al., 2018b) had a negative offset, i.e.: $BPF = 0.45\ MVF - 0.086$. But, this is assumed to be in error since it must be positive to describe the black curve.

[6] This linear equation was generated from the normalized water content estimated from $MWI$ reported in Figure 8 (West et al., 2018b) and the conversion to $MTV_{MWI}$ was done according to (Berman et al., 2018).

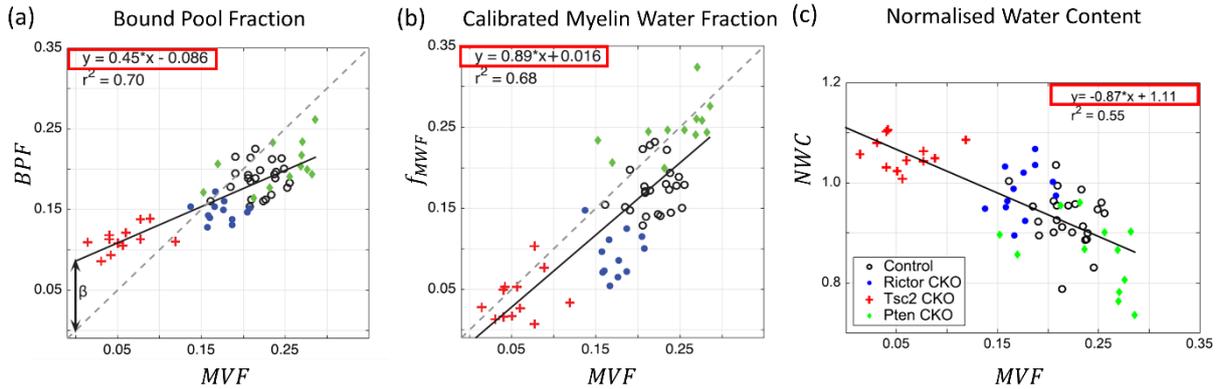

Adapted from West et al., 2018

**Figure 4:** Depicted are the linear relations between the myelin-volume fraction (MVF) from gold standard electron microscopy and three MRI-based biomarkers for myelin: (a) Bound Pool Fraction ($BPF$) from quantitative magnetization transfer imaging, (b) calibrated Myelinated Water Fraction ($f_{MWF}$) from myelin water imaging, (c) the Normalized Water Content ($NWC$). The estimated linear relations (red boxes) are used in our simulation experiment (see section 3.1.1-3.1.2). Macromolecular Tissue Volume ($MTV$) was calculated from $NWC$ according to Berman et al. (Berman et al., 2018). Modified and reproduced with permission from West et al. (West et al., 2018b).

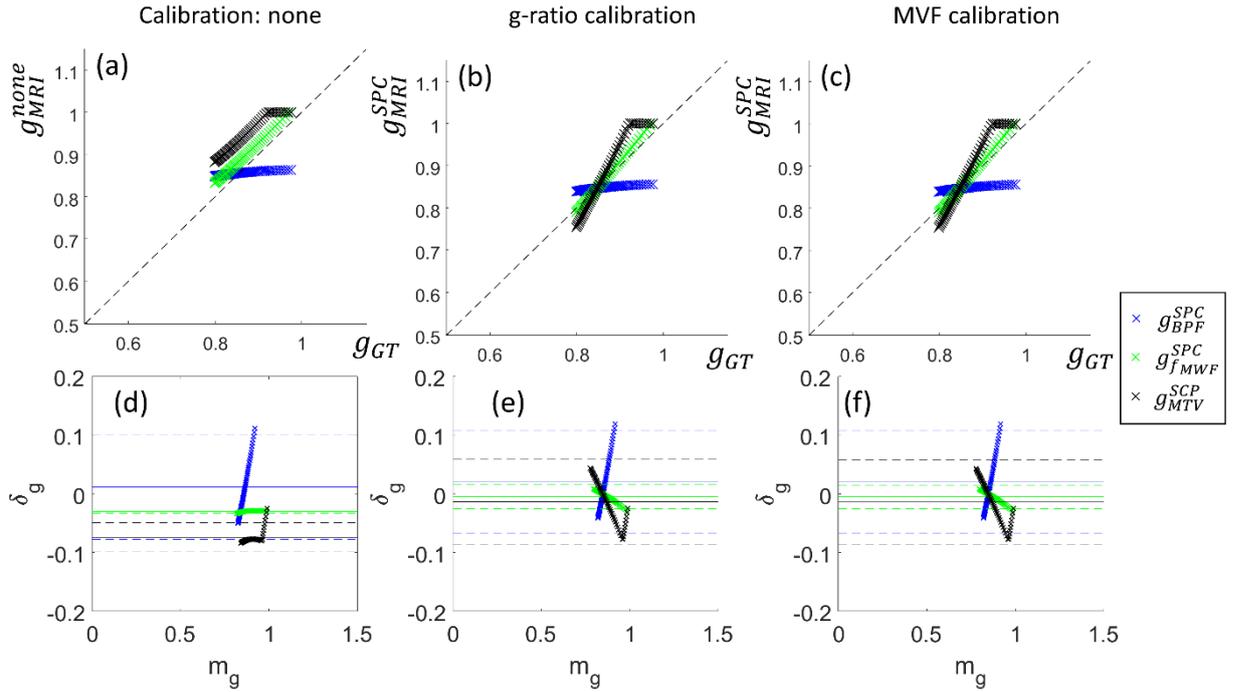

**Figure 5**: Depicted are scatter (a-c) and Bland-Altmann plots (d-f) of the ground truth ($g_{GT}$) and the MRI-based g-ratios using no calibration (a,d) or the single-point calibration approaches based on the g-ratio (b,e) or MVF (c,f) from a reference region of interest. The MRI-based g-ratios were calculated using different biomarkers for myelin: bound pool fraction ($BPF$, blue crosses), calibrated myelin-water fraction ($f_{MWF}$, green crosses), and macromolecular tissue volume ($MTV$, black crosses). The Bland-Altman plots (Bland and Altman, 1986) assess the bias and error when seeking to replace the ground truth g-ratio with the MRI-based measures. The plots depict the difference ($\delta_g = g_{GT} - g_{MRI}$) against the mean ($m_g = (g_{GT} + g_{MRI})/2$) of g-ratios with the solid line indicating the mean difference $\langle \delta_g \rangle$, and the dashed lines indicating $\langle \delta_g \rangle$ plus/minus 1.96 times the standard deviation of the differences $\langle std_{\delta_g} \rangle$.

### 3.2.2 Simulation results

The results are summarized in Figures 5 as well as in Table 2. Without calibration, we found that the bias was smallest for the BPF-based g-ratio (6.6%) and largest for the MTV-based g-ratio (-41.7%). The error was smallest for the MWF-based g-ratio (1.0%), moderate for the MTV-based g-ratio (7.1%) and largest for the BPF-based g-ratio (25.5%). Regardless of calibration method (i.e. MVF or g-ratio reference) the calibration reduced the bias for the MWF-based g-ratio (ca. -3%) and the MTV-based g-ratio (ca. -8%) but increased it for the BPF-based g-ratio (ca. 11%). Importantly, the calibration increased the error for MWF-based (ca. 6%) and MTV-based (ca. 21%) g-ratios and had almost no effect on the BPF-based g-ratio. Altogether, the BPF-based g-ratio exhibited comparatively large error exacerbated by its significantly compressed dynamic range ($dyn_{g_{BPF}} = \frac{\max(g_{BPF}^{none}) - \min(g_{BPF}^{none})}{g_{REF}} \approx 2\%$) as compared to the g-ratios using the other MVF proxies: $dyn_{g_{MWF}} \approx 25\%$, $dyn_{g_{MTV}} \approx 29\%$, $dyn_{g_{GT}} \approx 21\%$. Note that the abrupt change in the slopes of the MTV-based g-ratio values in Figs. 5 (black crosses) were due to reaching the upper limit for the MR g-ratio (i.e. the calibration led to a sealing effect for this biomarker).

| Calibration | BPF | | $f_{MWF}$ | | MTV | | Reference value |
|---|---|---|---|---|---|---|---|
| | Bias [%] | Error [%] | Bias [%] | Error [%] | Bias [%] | Error [%] | |
| none | 6.6 | 25.5 | −16.7 | 1.0 | −41.7 | 7.1 | − |
| g-ratio | 11.2 | 25.2 | −2.7 | 5.9 | −7.7 | 20.7 | 0.85 |
| MVF | 11.3 | 25.2 | −2.7 | 5.8 | −7.8 | 20.7 | 0.175 |

**Table 2**: The relative bias and error introduced by the g-ratio-based (3rd row) and MVF-based (4th row) single-point calibration as assessed by the Bland-Altman analysis for three different myelin biomarkers: Bound Pool Fraction ($BPF$ in 2nd column), calibrated Myelin Water Fraction ($f_{MWF}$ in 3rd column), and Macromolecular Tissue Volume ($MTV$ in 4th column). The reference values for the single-point calibrations are depicted in the last column. The bias and error are defined via the respective Bland-Altman analysis illustrated in Fig. 5. Bias is defined as the mean difference $\langle \delta_g \rangle$. Error is defined as the interval between +/-1.96 $\langle std_{\delta_g} \rangle$ with $\delta_g = g_{GT} - g_{MRI}$. Here, bias and error are presented in percentage, relative to the dynamic range of the ground truth g-ratios: $g_{dyn} = \max(g_{GT}) - \min(g_{GT}) = 0.18$.

### 3.2.3 What we can learn from the simulation experiment

We learned from this simulation that the single-point calibration can reduce the bias in the g-ratio (i.e. two out of three MRI-based g-ratio values became closer to the ground truth) but it comes at the cost of an increased error (i.e. the deviation from a one-to-one correspondence between the MR and the ground truth g-ratio increased after calibration). We expect that the latter feature is of more relevance to typical g-ratio studies, where longitudinal changes in the g-ratio or changes in the g-ratio between groups will likely be investigated. Moreover, the simulations showed that $f_{MWF}$ and $MTV_{MWI}$ are reasonable biomarkers for the g-ratio in terms of their error. Perhaps surprisingly, they perform best, in terms of error, when no calibration is performed. $BPF$, on the other hand, is degraded as an MVF biomarker when using single-point calibration, but also suffers from larger error when no calibration is performed. Interestingly, the two better performing MVF biomarkers, i.e. $f_{MWF}$ and $MTV_{MWI}$, both involved a calibration step in their computation, unlike the $BPF$. For $f_{MWF}$ the calibration was purely based on literature values, whereas $MTV_{MWI}$ was calibrated against a grey matter value specific to each brain.

Based on these simulations, a number of conclusions can be drawn. First, the single-point calibration method is insufficient to calibrate the g-ratio for the investigated scenarios where the offset parameter was non-zero. Second, at least in contexts consistent with those investigated here (e.g. fixed tissue, ex vivo MRI, mice), $BPF$-based g-ratio should not be used without more sophisticated calibration methods that are capable of accurately estimating both the slope and offset (see, e.g., (West et al., 2018b)). Third, $f_{MWF}$ and $MTV_{MWI}$ might be better biomarkers for g-ratio weighted imaging because they can be readily used without calibration. Some important caveats to these conclusions are outlined in the following.

Since gold standard information is missing when investigating *in vivo* data, we have used *ex vivo* data to generate ground truth data for a simulation experiment. It is important to bear in mind that the presented results may not translate to the *in vivo* case because of potentially different model validity (see section 3.1.1). Data quality can also vary considerably between *in vivo* and *ex vivo* imaging scenarios. Not only because of the use of fixed tissue *ex vivo* (and concomitant mitigation of physiological and motion corruption as well as the capacity for markedly longer scanning protocols) but also because of the different MRI techniques and non-clinical imaging systems used (West et al., 2018b). As a consequence, the relative differences in performance and impact of calibration for each of the g-ratios derived from different myelin biomarkers ($BPF$, $f_{MWF}$, and $MTV_{MWF}$) should be interpreted with care. This important point is further emphasised in figure 6. This demonstrates that the small dynamic range in the $BPF$-based g-ratio relative to the $MTV_{MWF}$-based g-ratio predicted by simulation (Fig. 6a), does not manifest in comparable *in vivo* g-ratio maps ($MT_{sat}$-based vs. $MTV$-based in Fig. 6b). These in fact show a greater dynamic range for the $MT_{sat}$-based g-ratio and higher correspondence between the two approaches after calibration. This contrasting observation might be due to the reasons outlined above, the use of somewhat different techniques *in vivo* and *ex vivo*, or due to fixation issues, e.g. fixation has been shown to strongly increase the $BPF$ in normal appearing white matter (Schmierer et al., 2008).

In summary, these simulations show that the single-point calibration, used in virtually all in vivo g-ratio mapping studies to date (Table 1), does not fully resolve the issue of converting MR proxies to the true MVF and can even increase bias and error in the g-ratio estimates. Therefore, further methodological development and validation is required to find the optimal means of ensuring the necessary validity and sensitivity of the MR g-ratio.

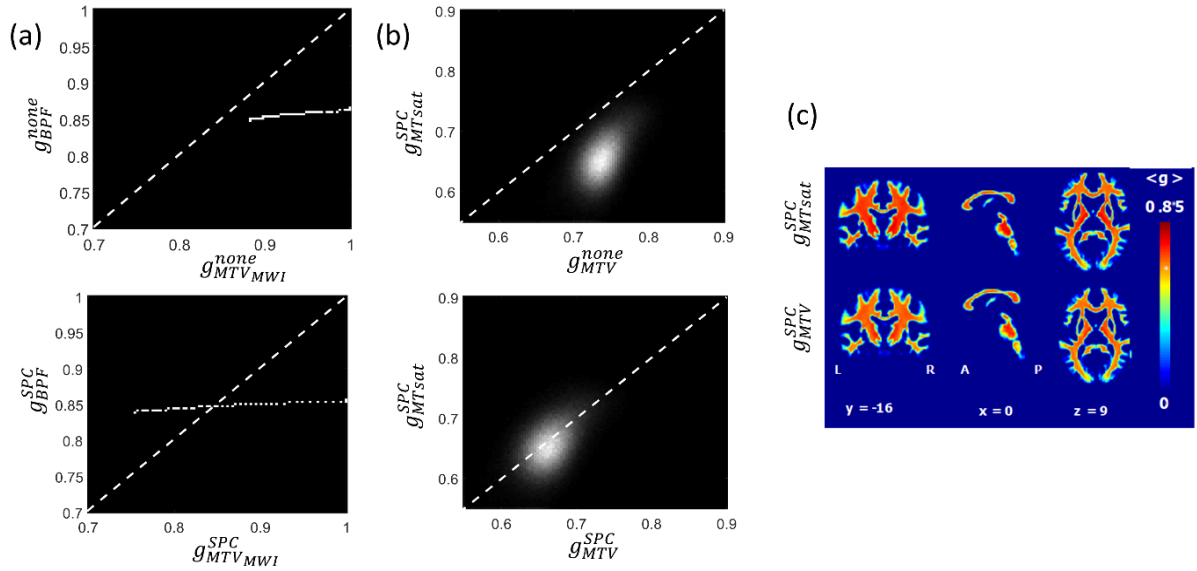

**Figure 6**: Illustration of the inter-relation between MR g-ratios derived from magnetisation transfer imaging (a: simulated $g_{BPF}$) and (b,c: in vivo $g_{MT_{sat}}$) or from the macromolecular tissue volume (a: simulated $g_{MTV_{MWI}}$) and (b,c: in vivo $g_{MTV}$) for two scenarios: omitting ($g^{none}$) or using single-point calibration ($g^{SPC}$) with a reference of $g_{REF} = 0.71$ in the medullary pyramid, estimated from (Graf von Keyserlingk and Schramm, 1984). The dynamic ranges of the MR g-ratios observed in simulation *ex vivo* (a) or via *in vivo* measurement (b,c) are very different. In both cases, there is a shift towards the identity line after SPC, but with much greater agreement between the measures *in vivo*. The maps in subfigure (c), adapted with permission from (Ellerbrock and Mohammadi, 2018a), were acquired using the protocol described in the caption of Fig. 7. Note that the MR g-ratios ("g3" and "g4") in the original publication were erroneous due to a reported mistake, see corrigendum (Ellerbrock and Mohammadi, 2018b). Here, the correct maps are depicted.

*3.3 Unification of Multi-modal data*

The aggregated g-ratio weighted imaging approach combines two complementary MRI contrasts, sensitive to the axonal-water and myelin volume fractions respectively. Given that each quantitative MRI technique is typically vulnerable to a specific set of artefacts, the combination of multiple data types needs to take care not to amplify these artefacts such that they obscure or corrupt the quantity of interest. For example, we have previously demonstrated that modality-specific spatial distortions, arising from inhomogeneous magnetic susceptibility distributions in the brain, can prevent voxel-wise spatial correspondence of the AWF and MVF proxies being achieved and lead to erroneous g-ratio estimates (Mohammadi et al., 2015). Even after correcting the susceptibility-induced distortions using dedicated tools (Ruthotto et al., 2012, 2013), residual misalignments between the EPI-based diffusion data and the gradient-echo based magnetisation transfer saturation maps can persist. The most obvious reason for residual misalignments is, of course, insufficient susceptibility distortion correction, but partial-volume effects in the EPI-based diffusion data associated with the typically lower spatial resolution, the EPI-readout, and eddy current distortions can also lead to lower white-matter tissue probability in the diffusion data relative to the MTsat map (Fig. 7). Here, we suggest combining the overlap between two modality-specific white-matter tissue probability maps (TPMs) to remove regions in the resulting g-ratio maps (Fig. 7a.v and 7b.v) that do not overlap between the two MRI contrasts, i.e. the region outside the red contours in Fig. 7a.iii and Fig. 7b.iii. In the example of Figure 7, the TPM was generated from the MTsat (Fig. 7a.iii and 7b.iii) and NODDI (Fig. 7a.iv and 7b.iv) map, respectively.

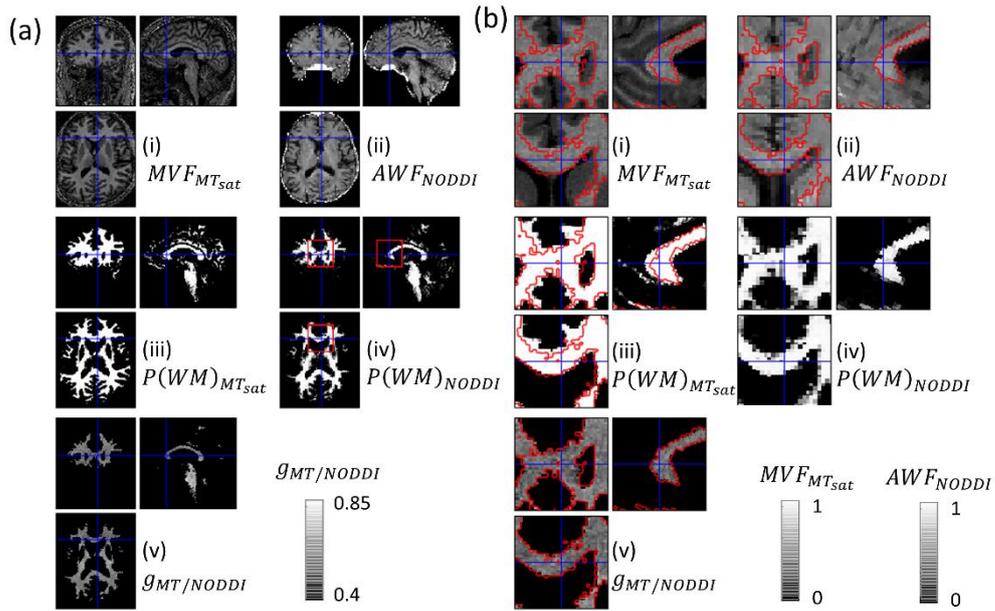

**Figure 7**: Depicted are whole brain views (a) (and magnifications (b)) of the aggregated g-ratio weighted map ((v): $g_{MT/NODDI}$), its constituting two qMRI maps: the calibrated myelin biomarker ((i): $MT_{sat}$), the axonal-water fraction ((ii): $AWF$), as well as the associated white-matter tissue probability maps (P(WM)) based on $MT_{sat}$ ((iii): $P(WM)_{MT_{sat}}$) or $AWF$ ((iv): $P(WM)_{AWF}$). Although, a distortion-corrected $AWF$ maps with negligible residual distortions was used, $P(WM)_{AWF}$ is lower than the $P(WM)_{MT_{sat}}$ (see contours in the magnification (b)). To reduce artefactual g-ratio values in regions where one of the two constituent biomarkers is not defined, we suggest generating g-ratio values only in voxels where both tissue probabilities exceed a pre-defined threshold (here 0.5). <u>Protocol details</u>: $MT_{sat}$: the MPM (Weiskopf et al., 2013) protocol, including calibration (Lutti et al., 2012) and three different weightings (MT-, PD- and $T_1$-weighting) SPGR images. The following sequence parameters were used: isotropic resolution of (1 mm)³, flip angle (FA) of 6° (for MT- and PD-weighted) and 21° (for T1-weighted), 8 echoes for PD- and T1-weighted images (2.3 to 18.4 ms, in steps of 2.3 ms) and 6 echoes for MT-weighted images (2.3 to 13.8 ms, in steps of 2.3 ms), readout bandwidth of 488 Hz/pixel, and repetition time (TR) of 25.00 ms. Two successive runs of the above mentioned SPGR data were acquired using a 3D acceleration factor of 3 and partial Fourier (6/8) (total time: 28 min.); $AWF$: 3 shell DWI (b-values:

500, 1000, 2500 mm/s2), each shell with 60 directions and additional b=0 images interspearsed, every 10 images, resolution: 1.6 mm isotropic, repetition time: 5300 ms, echo time: 73 ms. All scans were performed on a 3T PRISMA MRI (Siemens Healthcare, Erlangen, Germany), using the Siemens 1-channel transmitter (Tx) / 64-channel receiver (Rx) head-coil. The spatial distortions were reduced using the ACID toolbox (www.diffusiontools.com).

*3.4 Validation of g-ratio mapping*

Clear in vivo validation of g-ratio mapping is highly desirable, but generally unfeasible. We therefore typically rely on *ex vivo* histology for validation. A number of differences between these two imaging scenarios have been highlighted in previous sections. Here we summarise key points specifically pertinent to the *ex vivo* histology (e.g. electron microscopy) gold standard scenario. Most importantly, one has to consider the change in tissue composition that occurs when going from the **in vivo to the ex vivo situation**. In this case, the MRI signal and its parameters can significantly change due to, e.g. (i) autolysis (varying post-mortem interval (Shepherd et al., 2009)), (ii) fixation and the associated changes of cross-linking proteins, tissue shrinkage, and slowed diffusion processes (Schmierer et al., 2008; Shepherd et al., 2009), and (iii) temperature changes (Birkl et al., 2016). These changes affect diffusion (Dyrby et al., 2011) and other important MR parameters, such T1, T2* (Streubel et al., 2019) and susceptibility (based on signal phase) contrasts. However, despite these changes, the most important MRI mechanisms (e.g. diffusion anisotropy and relaxation mechanisms) are still present after fixation (Roebroeck et al., 2008). Nonetheless, it is necessary to characterize these differences in MRI parameters to enable translation and interpretation across *in vivo* and *ex vivo* measurements.

*3.4.1 g-ratio*

To date, only two studies have compared g-ratio measurements from ex vivo histology with MRI (Stikov et al., 2015; West et al., 2018a). Stikov et al. (2015) compared the g-ratio measured with *in* vivo MRI and *ex vivo* histology on a macaque monkey. West et al. (West et al., 2018a) compared g-ratio maps based on the WMTI, mcSMT and NODDI models to the equivalent g-ratio measured using gold standard histology techniques in mouse models. All three methods showed a moderate linear correspondence. It is important to note that for the NODDI model to work, the fixed diffusivities had to be adjusted empirically, suggesting that it is less well suited, at least for *ex vivo* data. Another interesting finding was that a simplified g-ratio model, in which the extra-axonal volume fraction was assumed to be one, such that AVF = 1 – MVF, performed equally well to the above mentioned diffusion signal models. The conclusions from this finding could be quite radical, i.e. that it is not necessary to measure both diffusion MRI and myelin markers to estimate changes in g-ratio across a strongly myelinating process. However, again caution is required since the gold-standard g-ratio (measured by histology) did not account for the contribution of unmyelinated axons, which the MRI g-ratio is also expected to depend on. Finally, it is important to highlight that, to date, no human specimen has been used to validate the g-ratio. This, however, would be a crucial step in linking ex vivo histology with our target in vivo application, i.e. g-ratio mapping in the human brain.

*3.4.2 MVF*

Here we discuss comparisons between myelin-sensitive MRI-based metrics and the gold standard MVF measured via histology that have been carried out in the context of g-ratio mapping. In early work, Stikov et al. compared the PSR estimated via MRI with the MVF estimated from electron microscopy (EM) in the corpus callosum of a macaque (Stikov et al., 2015). They found a linear dependence between these quantities but did not find the relationship to be significant, perhaps due to limited myelin-related variance present in the data. Using mouse models spanning hypo- and hyper-myelinated conditions has allowed a broader variance in myelination to be investigated (West et al., 2018b). West et al. used this approach to explore the relationship between the histological MVF, again derived from electron microscopy, and MRI-based measures in the same animals made using both MWI and qMT techniques (West et al., 2018b). They demonstrated a linear correlation between the MVF and both the MWF ($r = 0.81$) and the BPF ($r = 0.84$). Similar correlations were obtained between the MVF and the MR-derived equivalent, though the exact degree of correlation depended on the details of the calibration (Jung et al., 2018; West et al., 2018b). Berman et al. (Berman et al., 2018) used data from the same study to explore the dependence of the MTV on MVF. Unlike the typical quantification approach used *in vivo*, this *ex vivo* MTV measure was derived from a PD estimate obtained by extrapolating the MWI data to a TE of 0ms. A linear dependence on MVF was also demonstrated for this *ex vivo* MTV measure ($R^2 = 0.74$). While these *ex vivo* observations of linear dependence of myelin-sensitive MR metrics on MVF lend credence to the calibration approach investigated in section 3.1, they nonetheless reinforce the need for calibration since none show an offset-free 1:1 relationship. The previously outlined caveats regarding the translation of the methods from *ex vivo* to *in vivo* experiments must also be borne in mind.

3.4.3  *AVF*

Validation of AVF presents some distinct challenges. The AVF estimate derived from diffusion-based metrics (either via FVF or AVF=(1-MVF)AWF) is sensitive to the pool of myelinated axons but also influenced by the unmyelinated axons (Beaulieu and Allen, 1994a, 1994b; Beaulieu, 2009; Jones, 2010).  By contrast, gold standard EM-based assessment of volume fractions often focus on the myelinated axons only (Kelm et al., 2016; West et al., 2018b, 2018a; Zaimi et al., 2018; Tabarin et al., 2019). The myelin sheath provides protection against autolysis and acts as a contrast-enhancer for microscopy, making myelinated axons likely to be present and more easily detectable than unmyelinated axons (Olivares et al., 2001). In 2D EM, unmyelinated axons can also be confused with non-neuronal processes from cells like astrocytes or microglia. Ideally, a high-resolution microscopy approach combined with a neuron-specific stain, e.g. for neurofilaments, should be used to assess the AVF by encompassing all axons. (Jelescu et al., 2016b) compared MRI-based AWF with a histological counterpart (via Eq. (2)), including both myelinated axons and an estimate of unmyelinated axons, in mice models with different degrees of myelination. They found a linear relation, though not a 1:1 correspondence. This is an indication that MRI-based AWF also needs to be calibrated. However, it is expected that miscalibration of AWF will have less effect on MR g-ratio than miscalibration of MVF because it has been shown that (de)myelination-related changes in the g-ratio are strongly driven by changes in MVF and less so by the AWF (West et al., 2018a).

## 4    Conclusion and Outlook

This review provides methodological background for the MRI techniques pertinent to aggregate g-ratio weighted mapping with the aim of improving understanding of the currently used biomarkers, as well as to provide insight into the potentials and particularly the pitfalls. G-ratio weighted mapping has the potential to achieve non-invasive mapping of this functionally-relevant microstructural parameter by utilising the strength of multi-contrast quantitative MRI and biophysical models (also known as *in vivo* histology using MRI (Weiskopf et al., 2015)). The main take-home messages of this review are: (1) to fully benefit from the advantages of the aggregated g-ratio model, further work on a more appropriate calibration method is necessary to enable simultaneous estimation of both the slope and offset of the relationship between MRI markers and the true MVF; (2) more *ex vivo* histology gold standard measurements of human brain tissue are required to assess the typical range of MR g-ratio values that can be expected *in vivo*, (3) the quest to find the most appropriate MRI biomarkers for MVF and AVF for the *in vivo* situation is ongoing. In particular, there is currently a lack of validation studies for biomarkers of the AVF compartment using diffusion-based metrics. A major challenge here will be the estimation of the contribution to the AVF from unmyelinated axons (and cells potentially as well) via histology.

Other models that combine WMTI parameters and fibre dispersion (as defined by Watson distribution, e.g., in NODDI) (Jelescu et al., 2015; Jespersen et al., 2018) might have the potential to combine the sensitivity of WMTI to compartmental diffusivities with the less strict assumption about fibre alignment of the NODDI model. However, they suffer from model-inherent degeneracies (Jelescu et al., 2016a). One proposed solution to this degeneracy is to combine linear encoding schemes with planar or spherical diffusion sequences (Reisert et al., 2018; Coelho et al., 2019). A few studies have compared the diffusion anisotropy and intra-cellular signal fraction from linear diffusion weighting with planar diffusion weighting sequences: (Henriques et al., 2019) did it *ex vivo* in mice and (Mohammadi et al., 2017) did it *in vivo* in humans. However, these techniques have not yet been used for aggregated g-ratio weighted imaging. Another study has revealed a one-to-one correspondence between a simplified NODDI model and the mean diffusivity and fractional anisotropy as measured with DTI (Edwards et al., 2017). NODDI-DTI might help to link the models of g-ratio mapping studies based on a standard DTI protocol to those models that were based on more advanced diffusion MRI protocols. However, it has also not yet been applied to g-ratio mapping either. Future directions might also include the use of generative signal models that directly depend on the MR g-ratio (e.g., (Wharton and Bowtell, 2012, 2013)) to allow its extraction, or alternatively estimating the g-ratio from a multi-compartment GRE signal model (Thapaliya et al., 2018, 2020). New approaches that promise greater specificity to myelin (e.g. ihMT (Varma et al., 2015; Ercan et al., 2018; Duhamel et al., 2019)) and intra-axonal (Shemesh et al., 2016) compartments may also improve our capacity to directly map the g-ratio in the human brain *in vivo*.

## 5 Acknowledgements

We would like to thank Nadège Corbin, Luke J. Edwards, Francisco J. Fritz, Markus Morawski, Sebastian Papazoglou, and Nikolaus Weiskopf for helpful feedback and insightful discussion about the biophysical models. This work was supported by the German Research Foundation (DFG Priority Program 2041 "Computational Connectomics", [AL 1156/2-1;GE 2967/1-1; MO 2397/5-1; MO 2249/3–1], by the Emmy Noether Stipend: MO 2397/4-1), by the fmthh (01fmthh2017), and by the BMBF (01EW1711A and B) in the framework of ERA-NET NEURON. MFC is supported by the MRC and Spinal Research Charity through the ERA-NET Neuron joint call (MR/R000050/1). The Wellcome Centre for Human Neuroimaging is supported by core funding from the Wellcome [203147/Z/16/Z].

|       | Limitations of the AWF estimation approach (LA) |
|-------|---|
| LA.1  | Assumes parallel fibres and thus can be applied only in regions where this assumption is not violated (typically it has been applied in the corpus callosum). |
| LA.2  | The TFD (Reisert et al., 2012) has been assumed to be proportional to the fibre volume fraction, neglecting the contribution of the myelin water. It relies on a tractography algorithm and thus inherits the associated limitations. G-ratios based on this method show a larger scan-rescan variability as compared to NODDI-based g-ratios (Ellerbrock and Mohammadi, 2018a). |
| LA.3  | NODDI and mcSMT relate the perpendicular extra-axonal diffusivity to the parallel extra-axonal diffusivity scaled by "one minus the neurite density": ($D_{E,\perp} = (1-\nu)D_{E,\parallel}$). Moreover, NODDI and mcSMT impose a one-to-one scaling between the intra- and extra-cellular parallel diffusivities: $D_{A,\parallel} = D_{E,\parallel}$. |
| LA.4  | NODDI fixes it to a constant value (for *in vivo* healthy adults the diffusivities are usually assumed to be: $D_{A,\parallel} = D_{E,\parallel} = \frac{1.7 \mu m^2}{ms}$ and $D_0 = \frac{3 \mu m^2}{ms}$). |
| LA.5  | The WMTI model assumes parallel fibres and thus can applied only in regions where this assumption is not violated (typically it has been applied in the corpus callosum, but whether the model assumptions are sufficiently met there is unclear). |
| LA.6  | These studies provided not sufficient information to assess the specific implementation of the diffusion model. |
|       | **Limitation of the MVF estimation approach (LM)** |
| LM.1  | Requires a conversion factor to convert MRI-based myelin marker to the myelin volume fraction, which is done via histological data in different species. If this conversion factor is incorrect, the g-ratio will not be decoupled from the FVF (Stikov et al. 2015). |
| LM.2  | MTsat depends not only on the bound pool fraction but also on the rate of exchange, k, between the bound and free pools. Moreover it is a semi-quantitative measure because it depends on the particular off-resonance pulse used in the sequence, most notably its power and offset frequency (Helms et al., 2008). |
| LM.3  | Results in biased (over-)estimates (West et al. 2019) with artifactually high precision (West et al. 2019; Lankford and Does, 2013). |
| LM.4  | Requires estimation of the proton density and therefore a normalisation factor, e.g. the proton density in CSF. The optimal choice of the normalisation region will depend on the acquisition scheme, and will dictate the precision and accuracy of the MTV estimate. The modulation of the receiver coil's sensitivity also needs to be removed, either by constrained model fitting or measurement (Mezer et al. 2016). |
| LM.5  | The MTR suffers from the same limitations as MTsat, but retains dependence on both T1 and transmit-field inhomogeneities. These additional dependencies make it more prone to error as demonstrated e.g. in (Callaghan et al., 2015). |
| LM.6  | When fitting magnitude multi-echo data with long echo times, significant biases can be introduced by the Rician noise distribution that can greatly alter the measured $T_2(*)$ values (Bjarnason et al. 2013).<br>Fitting results are sensitive to the choice of TE and echo spacing, e.g. higher apparent $T_2(*)$ and smaller fractional contributions from short $T_2$ species as |
| | the first echo is increased (Whittall et al., 1999; Cercignani et al., 2018). A broad range of echo times are required to fully characterise both long and short T2 components.<br>Short echo times are required to acquire a signal with appreciable contribution from myelin, which is particularly problematic for gradient echo imaging due to the very short $T_2^*$ of myelin. |
| LM.7 | Error can result from the sensitivity to B1+ effects, both inhomogeneity, which can lead to stimulated echoes distorting the decay, and slice profile effects for 2D acquisitions (Lebel and Wilman, 2010). Power deposition can also be problematic, particularly at UHF. |
| LM.8 | Sensitivity to B0 inhomogeneity can bias model fits (Nam et al. 2015a). Phase errors caused by breathing and eddy currents can also lead to errors if uncorrected (Nam et al. 2015b). |
| LM.9 | Assumes a two pool model, which is a simplification, but likely sufficient to be supported by *in vivo* data acquired in the human brain (Levesque and Pike, 2009). |
| LM.10 | The model validity is unknown. |

TABLE A1: This table summarizes the limitation of MRI-based techniques for MVF and AWF (FVF) measurement.

| Table A2: MRI Methodological Abbreviations | |
|---|---|
| *Myelin Imaging Techniques* | |
| MWI | Myelin Water Imaging |
| MET2(*) | Multi-Exponential fitting to map compartment-specific $T_2(*)$ |
| qMT | Quantitative Magnetisation Transfer |
| bSSFP | Balanced Steady State Free Precession |
| SPGR | SPoiled Gradient Recalled echo |
| mcDESPOT | Multi-Compartment Driven Equilibrium Single Pulse Observation of $T_1$ and $T_2$ |
| | |
| *Diffusion Imaging Techniques* | |
| DWI | Diffusion Weighted Imaging |
| DTI | Diffusion Tensor Imaging |
| NODDI | Neurite Orientation and Dispersion Diffusing Imaging |
| CHARMED | Composite Hindered And Restricted Model of Diffusion |
| WMTI | White Matter Tissue Integrity |

| | | |
|---|---|---|
| TFD | Tract Fibre Density | |
| mcSMT | Multi-Compartment Spherical Mean Technique | |

*Biomarker and volume fractions*

| | | |
|---|---|---|
| BPF (f) | Bound pool fraction | Bound pool magnetisation relative to the combined bound and free pool magnetisation amplitudes as measured using qMT. |
| PSR (F) | Pool size ratio | Bound pool magnetisation relative to free pool magnetisation amplitude as measured using qMT. |
| AVF | Axonal volume fraction | The fraction of the imaging voxel volume that is intra-axonal. |
| AWF | Axonal water fraction | The fraction of the MRI water signal originating from the axonal compartment. |
| MVF | Myelin volume fraction | The fraction of the imaging voxel volume associated with myelin. This includes both the myelin itself and the water trapped between its bilayers. |
| MWF | Myelin water fraction | The fraction of the MRI water signal identified as exhibiting faster relaxation and attributed to the water trapped within the myelin sheath. |
| EVF | Extra cellular volume fraction | The fraction of the imaging voxel volume that originates outside the fibre. |
| FVF | Fibre volume fraction | The fraction of the imaging voxel volume that originates outside the fibre. |
| PD | Proton density | The concentration of MR-visible water relative to the concentration in the same volume comprised entirely of water. |
| MTV(F) | Macromolecular tissue volume (fraction) | The (fractional) volume of the imaging voxel that is comprised of macromolecules, i.e. that is not MR-visible water. |
| MTsat | Magnetisation transfer saturation | The steady state signal loss as a result of magnetisation transfer between the bound and free pools. |

# References

Alexander DC, Dyrby TB, Nilsson M, Zhang H (2019) Imaging brain microstructure with diffusion MRI: practicality and applications. NMR in Biomedicine 32:e3841.

Alexander DC, Hubbard PL, Hall MG, Moore EA, Ptito M, Parker GJM, Dyrby TB (2010) Orientationally invariant indices of axon diameter and density from diffusion MRI. Neuroimage 52:1374–1389.

Alonso-Ortiz E, Levesque IR, Pike GB (2015) MRI-based myelin water imaging: A technical review. Magnetic Resonance in Medicine 73:70–81.

Assaf Y, Basser PJ (2005) Composite hindered and restricted model of diffusion (CHARMED) MR imaging of the human brain. Neuroimage 27:48–58.

Assaf Y, Blumenfeld-Katzir T, Yovel Y, Basser PJ (2008) AxCaliber: a method for measuring axon diameter distribution from diffusion MRI. Magn Reson Med 59:1347–1354.

Assaf Y, Freidlin RZ, Rohde GK, Basser PJ (2004) New modeling and experimental framework to characterize hindered and restricted water diffusion in brain white matter. Magn Reson Med 52:965–978.

Baudrexel S, Reitz SC, Hof S, Gracien R-M, Fleischer V, Zimmermann H, Droby A, Klein JC, Deichmann R (2016) Quantitative T1 and proton density mapping with direct calculation of radiofrequency coil transmit and receive profiles from two-point variable flip angle data. NMR in Biomedicine 29:349–360.

Beaulieu C (2002) The basis of anisotropic water diffusion in the nervous system – a technical review. NMR in Biomedicine 15:435–455.

Beaulieu C (2009) CHAPTER 6 - The Biological Basis of Diffusion Anisotropy. In: Diffusion MRI (Johansen-Berg H, Behrens TEJ, eds), pp 105–126. San Diego: Academic Press. Available at: http://www.sciencedirect.com/science/article/pii/B9780123747099000067 [Accessed November 11, 2019].

Beaulieu C, Allen PS (1994a) Determinants of anisotropic water diffusion in nerves. Magn Reson Med 31:394–400.

Beaulieu C, Allen PS (1994b) Water diffusion in the giant axon of the squid: implications for diffusion-weighted MRI of the nervous system. Magn Reson Med 32:579–583.

Berman S, Filo S, Mezer AA (2019) Modeling conduction delays in the corpus callosum using MRI-measured g-ratio. Neuroimage 195:128–139.

Berman S, West KL, Does MD, Yeatman JD, Mezer AA (2018) Evaluating g-ratio weighted changes in the corpus callosum as a function of age and sex. Neuroimage 182:304–313.

Birkl C, Doucette J, Fan M, Hernandez-Torres E, Rauscher A (2020) Myelin water imaging depends on white matter fiber orientation in the human brain. bioRxiv:2020.03.11.987925.

Birkl C, Langkammer C, Golob-Schwarzl N, Leoni M, Haybaeck J, Goessler W, Fazekas F, Ropele S (2016) Effects of formalin fixation and temperature on MR relaxation times in the human brain. NMR Biomed 29:458–465.

Bjarnason TA, Laule C, Bluman J, Kozlowski P (2013) Temporal Phase Correction of Multiple Echo T2 Magnetic Resonance Images. J Magn Reson 231:22–31.

Bland JM, Altman DG (1986) Statistical methods for assessing agreement between two methods of clinical measurement. Lancet 1:307–310.

Cabana J-F, Gu Y, Boudreau M, Levesque IR, Atchia Y, Sled JG, Narayanan S, Arnold DL, Pike GB, Cohen-Adad J, Duval T, Vuong M-T, Stikov N (2015) Quantitative magnetization transfer imaging made easy with qMTLab: Software for data simulation, analysis, and visualization. Concepts in Magnetic Resonance Part A 44A:263–277.

Callaghan MF, Helms G, Lutti A, Mohammadi S, Weiskopf N (2015) A general linear relaxometry model of R1 using imaging data. Magnetic Resonance in Medicine 73:1309–1314.

Callaghan MF, Lutti A, Ashburner J, Balteau E, Corbin N, Draganski B, Helms G, Kherif F, Leutritz T, Mohammadi S (2019) Example dataset for the hMRI toolbox. Data in Brief:104132.

Campbell JSW, Leppert IR, Boudreau M, Narayanan S, Duval T, Cohen-Adad J, Pike GB, Stikov N (2018) Promise and pitfalls of g-ratio estimation with MRI. NeuroImage 182:80–96.

Cercignani M, Dowell NG, Tofts PS (2018) Quantitative MRI of the Brain: Principles of Physical Measurement, Second edition. CRC Press.

Cercignani M, Giulietti G, Dowell NG, Gabel M, Broad R, Leigh PN, Harrison NA, Bozzali M (2017) Characterizing axonal myelination within the healthy population: a tract-by-tract mapping of effects of age and gender on the fiber g-ratio. Neurobiol Aging 49:109–118.

Chomiak T, Hu B (2009) What is the optimal value of the g-ratio for myelinated fibers in the rat CNS? A theoretical approach. PLoS ONE 4:e7754.

Coelho S, Pozo JM, Jespersen SN, Jones DK, Frangi AF (2019) Resolving degeneracy in diffusion MRI biophysical model parameter estimation using double diffusion encoding. Magnetic Resonance in Medicine 82:395–410.

Coggan JS, Bittner S, Stiefel KM, Meuth SG, Prescott SA (2015) Physiological Dynamics in Demyelinating Diseases: Unraveling Complex Relationships through Computer Modeling. Int J Mol Sci 16:21215–21236.

David G, Freund P, Mohammadi S (2017) The efficiency of retrospective artifact correction methods in improving the statistical power of between-group differences in spinal cord DTI. NeuroImage 158:296–307.

Dean DC, O'Muircheartaigh J, Dirks H, Travers BG, Adluru N, Alexander AL, Deoni SCL (2016) Mapping an index of the myelin g-ratio in infants using magnetic resonance imaging. Neuroimage 132:225–237.

Deoni SCL, Matthews L, Kolind SH (2013) One component? Two components? Three? The effect of including a nonexchanging "free" water component in multicomponent driven equilibrium single pulse observation of T1 and T2. Magnetic Resonance in Medicine 70:147–154.

Deoni SCL, Rutt BK, Arun T, Pierpaoli C, Jones DK (2008) Gleaning multicomponent T1 and T2 information from steady-state imaging data. Magn Reson Med 60:1372–1387.

Does MD (2018) Inferring brain tissue composition and microstructure via MR relaxometry. Neuroimage.

Dortch RD, Harkins KD, Juttukonda MR, Gore JC, Does MD (2013) Characterizing inter-compartmental water exchange in myelinated tissue using relaxation exchange spectroscopy. Magnetic Resonance in Medicine 70:1450–1459.

Drakesmith M, Harms R, Rudrapatna SU, Parker GD, Evans CJ, Jones DK (2019) Estimating axon conduction velocity in vivo from microstructural MRI. Neuroimage 203:116186.

Duhamel G, Prevost VH, Cayre M, Hertanu A, Mchinda S, Carvalho VN, Varma G, Durbec P, Alsop DC, Girard OM (2019) Validating the sensitivity of inhomogeneous magnetization transfer (ihMT) MRI to myelin with fluorescence microscopy. NeuroImage 199:289–303.

Dula AN, Gochberg DF, Valentine HL, Valentine WM, Does MD (2010) Multiexponential T2, magnetization transfer, and quantitative histology in white matter tracts of rat spinal cord. Magn Reson Med 63:902–909.

Duval T, Lévy S, Stikov N, Campbell J, Mezer A, Witzel T, Keil B, Smith V, Wald LL, Klawiter E, Cohen-Adad J (2017) g-Ratio weighted imaging of the human spinal cord in vivo. Neuroimage 145:11–23.

Duval T, Smith V, Stikov N, Klawiter EC, Cohen-Adad J (2018) Scan-rescan of axcaliber, macromolecular tissue volume, and g-ratio in the spinal cord. Magn Reson Med 79:2759–2765.

Dyrby TB, Baaré WFC, Alexander DC, Jelsing J, Garde E, Søgaard LV (2011) An ex vivo imaging pipeline for producing high-quality and high-resolution diffusion-weighted imaging datasets. Hum Brain Mapp 32:544–563.

Edwards LJ, Pine KJ, Ellerbrock I, Weiskopf N, Mohammadi S (2017) NODDI-DTI: Estimating Neurite Orientation and Dispersion Parameters from a Diffusion Tensor in Healthy White Matter. Front Neurosci 11:720.

Ellerbrock I, Mohammadi S (2018a) Four in vivo g-ratio-weighted imaging methods: Comparability and repeatability at the group level. Hum Brain Mapp 39:24–41.

Ellerbrock I, Mohammadi S (2018b) Corrigendum to Ellerbrock et al. (2018) "Four in vivo g-ratio-weighted imaging methods: Comparability and repeatability at the group level." Human Brain Mapping 39:1467–1467.

Eng J, Ceckler TL, Balaban RS (1991) Quantitative 1H magnetization transfer imaging in vivo. Magn Reson Med 17:304–314.

Ercan E, Varma G, Mädler B, Dimitrov IE, Pinho MC, Xi Y, Wagner BC, Davenport EM, Maldjian JA, Alsop DC, Lenkinski RE, Vinogradov E (2018) Microstructural correlates of 3D steady-state inhomogeneous magnetization transfer (ihMT) in the human brain white matter assessed by myelin water imaging and diffusion tensor imaging. Magnetic Resonance in Medicine 80:2402–2414.

Fields RD (2015) A new mechanism of nervous system plasticity: activity-dependent myelination. Nature Reviews Neuroscience 16:756–767.

Fieremans E, Jensen JH, Helpern JA (2011) White matter characterization with diffusional kurtosis imaging. Neuroimage 58:177–188.

Graf von Keyserlingk D, Schramm U (1984) Diameter of axons and thickness of myelin sheaths of the pyramidal tract fibres in the adult human medullary pyramid. Anat Anz 157:97–111.

Hagiwara A, Hori M, Yokoyama K, Nakazawa M, Ueda R, Horita M, Andica C, Abe O, Aoki S (2017) Analysis of White Matter Damage in Patients with Multiple Sclerosis via a Novel In Vivo MR Method for Measuring Myelin, Axons, and G-Ratio. American Journal of Neuroradiology 38:1934–1940.

Harkins KD, Dula AN, Does MD (2012) Effect of intercompartmental water exchange on the apparent myelin water fraction in multiexponential T2 measurements of rat spinal cord. Magn Reson Med 67:793–800.

Hartline DK, Colman DR (2007) Rapid Conduction and the Evolution of Giant Axons and Myelinated Fibers. Current Biology 17:R29–R35.

Helms G, Dathe H, Kallenberg K, Dechent P (2008) High-resolution maps of magnetization transfer with inherent correction for RF inhomogeneity and T1 relaxation obtained from 3D FLASH MRI. Magn Reson Med 60:1396–1407.

Henkelman RM, Stanisz GJ, Graham SJ (2001) Magnetization transfer in MRI: a review. NMR Biomed 14:57–64.

Henriques RN, Jespersen SN, Shemesh N (2019) Microscopic anisotropy misestimation in spherical-mean single diffusion encoding MRI. Magnetic Resonance in Medicine 81:3245–3261.

Hildebrand C, Hahn R (1978) Relation between myelin sheath thickness and axon size in spinal cord white matter of some vertebrate species. J Neurol Sci 38:421–434.

Hori M, Hagiwara A, Fukunaga I, Ueda R, Kamiya K, Suzuki Y, Liu W, Murata K, Takamura T, Hamasaki N, Irie R, Kamagata K, Kumamaru KK, Suzuki M, Aoki S (2018) Application of Quantitative Microstructural MR Imaging with Atlas-based Analysis for the Spinal Cord in Cervical Spondylotic Myelopathy. Sci Rep 8 Available at: https://www.ncbi.nlm.nih.gov/pmc/articles/PMC5979956/ [Accessed February 18, 2020].

Huang SY, Tobyne SM, Nummenmaa A, Witzel T, Wald LL, McNab JA, Klawiter EC (2016) Characterization of Axonal Disease in Patients with Multiple Sclerosis Using High-Gradient-Diffusion MR Imaging. Radiology 280:244–251.

Jang H, Ma Y, Searleman AC, Carl M, Corey-Bloom J, Chang EY, Du J (2020) Inversion recovery UTE based volumetric myelin imaging in human brain using interleaved hybrid encoding. Magn Reson Med 83:950–961.

Jelescu IO, Budde MD (2017) Design and Validation of Diffusion MRI Models of White Matter. Front Phys 5 Available at: https://www.frontiersin.org/articles/10.3389/fphy.2017.00061/full#B56 [Accessed May 9, 2019].

Jelescu IO, Veraart J, Adisetiyo V, Milla SS, Novikov DS, Fieremans E (2015) One diffusion acquisition and different white matter models: how does microstructure change in human early development based on WMTI and NODDI? Neuroimage 107:242–256.

Jelescu IO, Veraart J, Fieremans E, Novikov DS (2016a) Degeneracy in model parameter estimation for multi-compartmental diffusion in neuronal tissue. NMR Biomed 29:33–47.

Jelescu IO, Zurek M, Winters KV, Veraart J, Rajaratnam A, Kim NS, Babb JS, Shepherd TM, Novikov DS, Kim SG, Fieremans E (2016b) In vivo quantification of demyelination and recovery using compartment-specific diffusion MRI metrics validated by electron microscopy. Neuroimage 132:104–114.

Jespersen SN, Leigland LA, Cornea A, Kroenke CD (2012) Determination of axonal and dendritic orientation distributions within the developing cerebral cortex by diffusion tensor imaging. IEEE Trans Med Imaging 31:16–32.

Jespersen SN, Olesen JL, Hansen B, Shemesh N (2018) Diffusion time dependence of microstructural parameters in fixed spinal cord. Neuroimage 182:329–342.

Jones DK (2010) Diffusion MRI: Theory, Methods, and Applications. Oxford University Press.

Jung W, Lee J, Shin H-G, Nam Y, Zhang H, Oh S-H, Lee J (2018) Whole brain g-ratio mapping using myelin water imaging (MWI) and neurite orientation dispersion and density imaging (NODDI). Neuroimage 182:379–388.

Kaden E, Kelm ND, Carson RP, Does MD, Alexander DC (2016) Multi-compartment microscopic diffusion imaging. Neuroimage 139:346–359.

Kamagata K, Zalesky A, Yokoyama K, Andica C, Hagiwara A, Shimoji K, Kumamaru KK, Takemura MY, Hoshino Y, Kamiya K, Hori M, Pantelis C, Hattori N, Aoki S (2019) MR g-ratio-weighted connectome analysis in patients with multiple sclerosis. Sci Rep 9:1–13.

Kelm ND, West KL, Carson RP, Gochberg DF, Ess KC, Does MD (2016) Evaluation of diffusion kurtosis imaging in ex vivo hypomyelinated mouse brains. Neuroimage 124:612–626.

Lankford CL, Does MD (2013) On the inherent precision of mcDESPOT. Magn Reson Med 69:127–136.

Lebel RM, Wilman AH (2010) Transverse relaxometry with stimulated echo compensation. Magn Reson Med 64:1005–1014.

Lenz C, Klarhöfer M, Scheffler K (2012) Feasibility of in vivo myelin water imaging using 3D multigradient-echo pulse sequences. Magn Reson Med 68:523–528.

Levesque IR, Pike GB (2009) Characterizing healthy and diseased white matter using quantitative magnetization transfer and multicomponent T2 relaxometry: A unified view via a four-pool model. Magnetic Resonance in Medicine 62:1487–1496.

Liu F, Block WF, Kijowski R, Samsonov A (2016) Rapid multicomponent relaxometry in steady state with correction of magnetization transfer effects. Magn Reson Med 75:1423–1433.

Lorio S, Tierney TM, McDowell A, Arthurs OJ, Lutti A, Weiskopf N, Carmichael DW (2019) Flexible proton density (PD) mapping using multi-contrast variable flip angle (VFA) data. Neuroimage 186:464–475.

Lutti A, Stadler J, Josephs O, Windischberger C, Speck O, Bernarding J, Hutton C, Weiskopf N (2012) Robust and fast whole brain mapping of the RF transmit field B1 at 7T. PLoS ONE 7:e32379.

MacKay A, Laule C, Vavasour I, Bjarnason T, Kolind S, Mädler B (2006) Insights into brain microstructure from the T2 distribution. Magnetic Resonance Imaging 24:515–525.

MacKay AL, Laule C (2016) Magnetic Resonance of Myelin Water: An in vivo Marker for Myelin Zalc B, ed. BPL 2:71–91.

Magerkurth J, Volz S, Wagner M, Jurcoane A, Anti S, Seiler A, Hattingen E, Deichmann R (2011) Quantitative T*2-mapping based on multi-slice multiple gradient echo flash imaging: retrospective correction for subject motion effects. Magn Reson Med 66:989–997.

Mancini M, Giulietti G, Dowell N, Spanò B, Harrison N, Bozzali M, Cercignani M (2018) Introducing axonal myelination in connectomics: A preliminary analysis of g-ratio distribution in healthy subjects. Neuroimage 182:351–359.

McConnell HM (1958) Reaction Rates by Nuclear Magnetic Resonance. J Chem Phys 28:430–431.

McKinnon ET, Jensen JH (2019) Measuring intra-axonal T2 in white matter with direction-averaged diffusion MRI. Magn Reson Med 81:2985–2994.

Melbourne A, Eaton-Rosen Z, Orasanu E, Price D, Bainbridge A, Cardoso MJ, Kendall GS, Robertson NJ, Marlow N, Ourselin S (2016) Longitudinal development in the preterm thalamus and posterior white matter: MRI correlations between diffusion weighted imaging and T2 relaxometry. Hum Brain Mapp 37:2479–2492.

Mezer A, Rokem A, Berman S, Hastie T, Wandell BA (2016) Evaluating quantitative proton-density-mapping methods. Human Brain Mapping 37:3623–3635.

Mezer A, Yeatman JD, Stikov N, Kay KN, Cho N-J, Dougherty RF, Perry ML, Parvizi J, Hua LH, Butts-Pauly K, Wandell BA (2013) Quantifying the local tissue volume and composition in individual brains with magnetic resonance imaging. Nat Med 19:1667–1672.

Mohammadi S, Carey D, Dick F, Diedrichsen J, Sereno MI, Reisert M, Callaghan MF, Weiskopf N (2015) Whole-brain in-vivo measurements of the axonal g-ratio in a group of 37 healthy volunteers. Frontiers in neuroscience 9:441.

Mohammadi S, Ellerbrock I, Edwards L (2017) Biomarkers for fiber density: comparing Stejskal-Tanner diffusion encoding metrics with microscopic diffusion anisotropy from double-diffusion encoding imaging. In: Proc. Intl. Soc. Mag. Reson. Med. 25 (2017), abstract: 3382.

Morrison C, Henkelman RM (1995) A Model for Magnetization Transfer in Tissues. Magnetic Resonance in Medicine 33:475–482.

Nam Y, Kim D-H, Lee J (2015a) Physiological noise compensation in gradient-echo myelin water imaging. NeuroImage 120:345–349.

Nam Y, Lee J, Hwang D, Kim D-H (2015b) Improved estimation of myelin water fraction using complex model fitting. NeuroImage 116:214–221.

Nöth U, Shrestha M, Schüre J-R, Deichmann R (2017) Quantitative in vivo T2 mapping using fast spin echo techniques - A linear correction procedure. Neuroimage 157:476–485.

Novikov DS, Fieremans E, Jespersen SN, Kiselev VG (2019) Quantifying brain microstructure with diffusion MRI: Theory and parameter estimation. NMR in Biomedicine 32:e3998.

Olivares R, Montiel J, Aboitiz F (2001) Species Differences and Similarities in the Fine Structure of the Mammalian Corpus callosum. BBE 57:98–105.

Reisert M, Kiselev VG, Dhital B (2018) A Unique Analytical Solution of the White Matter Standard Model using Linear and Planar Encodings. arXiv:180804389 [physics, q-bio] Available at: http://arxiv.org/abs/1808.04389 [Accessed May 30, 2019].

Reisert M, Mader I, Umarova R, Maier S, Tebartz van Elst L, Kiselev VG (2013) Fiber density estimation from single q-shell diffusion imaging by tensor divergence. Neuroimage 77:166–176.

Reisert M, Skibbe H, Kiselev VG (2012) Fiber density estimation by tensor divergence. Med Image Comput Comput Assist Interv 15:297–304.

Roebroeck A, Galuske R, Formisano E, Chiry O, Bratzke H, Ronen I, Kim D, Goebel R (2008) High-resolution diffusion tensor imaging and tractography of the human optic chiasm at 9.4 T. Neuroimage 39:157–168.

Rushton WAH (1951) A theory of the effects of fibre size in medullated nerve. J Physiol (Lond) 115:101–122.

Ruthotto L, Kugel H, Olesch J, Fischer B, Modersitzki J, Burger M, Wolters CH (2012) Diffeomorphic susceptibility artifact correction of diffusion-weighted magnetic resonance images. Phys Med Biol 57:5715–5731.

Ruthotto L, Mohammadi S, Heck C, Modersitzki J, Weiskopf N (2013) Hyperelastic Susceptibility Artifact Correction of DTI in SPM. In: Bildverarbeitung für die Medizin 2013, pp 344–349. Springer.

Salami M, Itami C, Tsumoto T, Kimura F (2003) Change of conduction velocity by regional myelination yields constant latency irrespective of distance between thalamus and cortex. Proc Natl Acad Sci USA 100:6174–6179.

Sati P, van Gelderen P, Silva AC, Reich DS, Merkle H, de Zwart JA, Duyn JH (2013) Micro-compartment specific T2* relaxation in the brain. Neuroimage 77:268–278.

Schmidt H, Knösche TR (2019) Action potential propagation and synchronisation in myelinated axons. PLOS Computational Biology 15:e1007004.

Schmierer K, Scaravilli F, Altmann DR, Barker GJ, Miller DH (2004) Magnetization transfer ratio and myelin in postmortem multiple sclerosis brain. Ann Neurol 56:407–415.

Schmierer K, Wheeler-Kingshott CA, Tozer DJ, Boulby PA, Parkes HG, Yousry TA, Scaravilli F, Barker GJ, Tofts PS, Miller DH (2008) Quantitative Magnetic Resonance of Post Mortem Multiple Sclerosis Brain before and after Fixation. Magn Reson Med 59:268–277.

Shemesh N, Jespersen SN, Alexander DC, Cohen Y, Drobnjak I, Dyrby TB, Finsterbusch J, Koch MA, Kuder T, Laun F, Lawrenz M, Lundell H, Mitra PP, Nilsson M, Özarslan E, Topgaard D, Westin C-F (2016) Conventions and nomenclature for double diffusion encoding NMR and MRI. Magn Reson Med 75:82–87.

Shepherd TM, Flint JJ, Thelwall PE, Stanisz GJ, Mareci TH, Yachnis AT, Blackband SJ (2009) Postmortem interval alters the water relaxation and diffusion properties of rat nervous tissue--implications for MRI studies of human autopsy samples. Neuroimage 44:820–826.

Sheth V, Shao H, Chen J, Vandenberg S, Corey-Bloom J, Bydder GM, Du J (2016) Magnetic resonance imaging of myelin using ultrashort Echo time (UTE) pulse sequences: Phantom, specimen, volunteer and multiple sclerosis patient studies. Neuroimage 136:37–44.

Sled JG (2018) Modelling and interpretation of magnetization transfer imaging in the brain. NeuroImage 182:128–135.

Sled JG, Levesque I, Santos AC, Francis SJ, Narayanan S, Brass SD, Arnold DL, Pike GB (2004) Regional variations in normal brain shown by quantitative magnetization transfer imaging. Magn Reson Med 51:299–303.

Sled JG, Pike GB (2001) Quantitative imaging of magnetization transfer exchange and relaxation properties in vivo using MRI. Magn Reson Med 46:923–931.

Stejskal EO, Tanner JE (1965) Spin Diffusion Measurements: Spin Echoes in the Presence of a Time-Dependent Field Gradient. The Journal of Chemical Physics 42:288.

Stikov N, Campbell JSW, Stroh T, Lavelée M, Frey S, Novek J, Nuara S, Ho M-K, Bedell BJ, Dougherty RF, Leppert IR, Boudreau M, Narayanan S, Duval T, Cohen-Adad J, Picard P-A, Gasecka A, Côté D, Pike GB (2015) In vivo histology of the myelin g-ratio with magnetic resonance imaging. NeuroImage 118:397–405.

Stikov N, Perry LM, Mezer A, Rykhlevskaia E, Wandell BA, Pauly JM, Dougherty RF (2011) Bound pool fractions complement diffusion measures to describe white matter micro and macrostructure. Neuroimage 54:1112–1121.

Stoyan D (1988) Fisher, N. I., T. Lewis and B. J. Jempleton: Statistical Analysis of Spherical Data. Cambridge University Press, Cambridge – New York – New Rochelle – Melbourne – Sydney 1987, XIV, 329 S., $ 65.—. Biometrical Journal 30:868–868.

Streubel S, Ashtarayeh M, Mushumba H, Papazoglou S, Püschel K, Streubel T (2019) Longitudinal assessment of relaxation and magnetization transfer saturation rates during formalin fixation across fiber pathways of the human brain. In: Proc Intl Soc Magn Reson Med. 2019;28: #1212.

Szafer A, Zhong J, Gore JC (1995) Theoretical model for water diffusion in tissues. Magn Reson Med 33:697–712.

Tabarin T, Morozova M, Jaeger C, Rush H, Morawski M, Geyer S, Mohammadi S (2019) Deep learning segmentation (AxonDeepSeg) to generate axonal-property map from ex vivo human optic chiasm using light microscopy. In: Proc Intl Soc Magn Reson Med. 2019;28: #4722.

Tabelow K, Balteau E, Ashburner J, Callaghan MF, Draganski B, Helms G, Kherif F, Leutritz T, Lutti A, Phillips C, Reimer E, Ruthotto L, Seif M, Weiskopf N, Ziegler G, Mohammadi S (2019) hMRI – A toolbox for quantitative MRI in neuroscience and clinical research. NeuroImage 194:191–210.

Thapaliya K, Vegh V, Bollmann S, Barth M (2018) Assessment of microstructural signal compartments across the corpus callosum using multi-echo gradient recalled echo at 7 T. Neuroimage 182:407–416.

Thapaliya K, Vegh V, Bollmann S, Barth M (2020) Influence of 7T GRE-MRI Signal Compartment Model Choice on Tissue Parameters. Front Neurosci 14 Available at: https://www.frontiersin.org/articles/10.3389/fnins.2020.00271/full [Accessed May 15, 2020].

Tofts P (2004) Quantitative MRI of the Brain: Measuring Changes Caused by Disease, 1. Auflage. John Wiley & Sons.

van Zijl PCM, Lam WW, Xu J, Knutsson L, Stanisz GJ (2018) Magnetization Transfer Contrast and Chemical Exchange Saturation Transfer MRI. Features and analysis of the field-dependent saturation spectrum. NeuroImage 168:222–241.

Varma G, Duhamel G, de Bazelaire C, Alsop DC (2015) Magnetization transfer from inhomogeneously broadened lines: A potential marker for myelin. Magn Reson Med 73:614–622.

Veraart J, Novikov DS, Fieremans E (2018) TE dependent Diffusion Imaging (TEdDI) distinguishes between compartmental T2 relaxation times. Neuroimage 182:360–369.

Volz S, Nöth U, Jurcoane A, Ziemann U, Hattingen E, Deichmann R (2012) Quantitative proton density mapping: correcting the receiver sensitivity bias via pseudo proton densities. Neuroimage 63:540–552.

Wang Y, Chen Y, Wu D, Wang Y, Sethi SK, Yang G, Xie H, Xia S, Haacke EM (2018) STrategically Acquired Gradient Echo (STAGE) imaging, part II: Correcting for RF inhomogeneities in estimating T1 and proton density. Magn Reson Imaging 46:140–150.

Warntjes JBM, Dahlqvist O, Lundberg P (2007) Novel method for rapid, simultaneous T1, T2*, and proton density quantification. Magn Reson Med 57:528–537.

Warntjes M, Engström M, Tisell A, Lundberg P (2016) Modeling the Presence of Myelin and Edema in the Brain Based on Multi-Parametric Quantitative MRI. Front Neurol 7 Available at: https://www.frontiersin.org/articles/10.3389/fneur.2016.00016/full [Accessed May 17, 2020].

Webb S, Munro CA, Midha R, Stanisz GJ (2003) Is multicomponent T2 a good measure of myelin content in peripheral nerve? Magn Reson Med 49:638–645.

Weiger M, Froidevaux R, Baadsvik EL, Brunner DO, Rösler MB, Pruessmann KP (2020) Advances in MRI of the myelin bilayer. NeuroImage 217:116888.

Weiskopf N, Mohammadi S, Lutti A, Callaghan MF (2015) Advances in MRI-based computational neuroanatomy: from morphometry to in-vivo histology. Current opinion in neurology 28:313–322.

Weiskopf N, Suckling J, Williams G, Correia MM, Inkster B, Tait R, Ooi C, Bullmore ET, Lutti A (2013) Quantitative multi-parameter mapping of R1, PD*, MT and R2* at 3T: a multi-center validation. Front Neurosci 7:95.

West DJ, Teixeira RPAG, Wood TC, Hajnal JV, Tournier J-D, Malik SJ (2019) Inherent and unpredictable bias in multi-component DESPOT myelin water fraction estimation. NeuroImage 195:78–88.

West KL, Kelm ND, Carson RP, Alexander DC, Gochberg DF, Does MD (2018a) Experimental studies of g-ratio MRI in ex vivo mouse brain. Neuroimage 167:366–371.

West KL, Kelm ND, Carson RP, Does MD (2016) A revised model for estimating g-ratio from MRI. Neuroimage 125:1155–1158.

West KL, Kelm ND, Carson RP, Gochberg DF, Ess KC, Does MD (2018b) Myelin volume fraction imaging with MRI. Neuroimage 182:511–521.

Wharton S, Bowtell R (2012) Fiber orientation-dependent white matter contrast in gradient echo MRI. PNAS 109:18559–18564.

Wharton S, Bowtell R (2013) Gradient echo based fiber orientation mapping using R2* and frequency difference measurements. NeuroImage 83:1011–1023.

Whittall KP, MacKay AL (1989) Quantitative interpretation of NMR relaxation data. Journal of Magnetic Resonance (1969) 84:134–152.

Whittall KP, Mackay AL, Graeb DA, Nugent RA, Li DKB, Paty DW (1997) In vivo measurement of T2 distributions and water contents in normal human brain. Magnetic Resonance in Medicine 37:34–43.

Whittall KP, MacKay AL, Li DKB (1999) Are mono-exponential fits to a few echoes sufficient to determine T2 relaxation for in vivo human brain? Magnetic Resonance in Medicine 41:1255–1257.

Wolff SD, Balaban RS (1989) Magnetization transfer contrast (MTC) and tissue water proton relaxation in vivo. Magnetic Resonance in Medicine 10:135–144.

Yiannakas MC, Kearney H, Samson RS, Chard DT, Ciccarelli O, Miller DH, Wheeler-Kingshott CAM (2012) Feasibility of grey matter and white matter segmentation of the upper cervical cord in vivo: A pilot study with application to magnetisation transfer measurements. NeuroImage 63:1054–1059.

Yu F, Fan Q, Tian Q, Ngamsombat C, Machado N, Bireley JD, Russo AW, Nummenmaa A, Witzel T, Wald LL, Klawiter EC, Huang SY (2019) Imaging G-Ratio in Multiple Sclerosis Using High-Gradient Diffusion MRI and Macromolecular Tissue Volume. AJNR Am J Neuroradiol 40:1871–1877.

Zaimi A, Wabartha M, Herman V, Antonsanti P-L, Perone CS, Cohen-Adad J (2018) AxonDeepSeg: automatic axon and myelin segmentation from microscopy data using convolutional neural networks. Sci Rep 8:3816.

Zhang H, Schneider T, Wheeler-Kingshott CA, Alexander DC (2012) NODDI: practical in vivo neurite orientation dispersion and density imaging of the human brain. Neuroimage 61:1000–1016.

Zimmerman JR, Brittin WE (1957) Nuclear Magnetic Resonance Studies in Multiple Phase Systems: Lifetime of a Water Molecule in an Adsorbing Phase on Silica Gel. J Phys Chem 61:1328–1333.